\documentclass{elsarticle}
\usepackage{graphicx}

\usepackage{times} 
\usepackage{amsmath}
\usepackage{amssymb}

\usepackage{color}

\usepackage{graphics} 
\usepackage{epsfig} 
\usepackage{subfigure}
\usepackage{ulem}

\usepackage{tikz}
\usetikzlibrary{patterns}
\usepackage{algorithmic}
\usepackage{algorithm}

\newcommand{\dt}[0]{\mbox{d}t}

\newcommand{\R}{\mathbb{R}}

\newcommand{\J}{\mathcal{J}}

\usepackage{bm}
\usepackage[colorlinks=true]{hyperref}
\hypersetup{urlcolor=blue, citecolor=red}

\newtheorem{theorem}{Theorem}[section]
\newtheorem{corollary}{Corollary}
\newtheorem{lemma}[theorem]{Lemma}
\newtheorem{proposition}{Proposition}

\newtheorem{definition}[theorem]{Definition}
\newtheorem{remark}{Remark}
\newtheorem{example}{Example}

\newcommand{\ra}{\rightarrow}

\newcommand{\F}{\mathbb{F}}

\usepackage{amssymb}

\newcommand{\proa}{A^*G \mbox{$\;$}_{\tau^*} \kern-3pt\times_\alpha
G \mbox{$\;$}_\beta \kern-3pt\times_{\tau^*} A^*G}

\tikzstyle{vertex}=[circle,fill=black!20,minimum size=15pt,inner sep=0pt]
\tikzstyle{selected vertex} = [vertex, fill=red!24]
\tikzstyle{edge} = [draw,thick,-]
\tikzstyle{dedge} = [draw,thick,<->]
\tikzstyle{shadowdedge} = [draw, dotted,->]
\tikzstyle{weight} = [font=\small]
\tikzstyle{selected edge} = [draw,line width=3pt,-,red!50]
\tikzstyle{ignored edge} = [draw,line width=3pt,-,black!20]

\begin{document}

\begin{frontmatter}

\title{Variational integrators for non-autonomous systems \\ with applications to stabilization of multi-agent formations}
\author[First]{Leonardo Colombo}
\author[Second]{Manuela Gamonal Fern\'andez}
\author[Second]{David Mart\'in de Diego}

\address[First]{Centre for Automation and Robotics
(CSIC-UPM). Ctra. M300 Campo Real, Km 0,200, Arganda
del Rey, 28500 Madrid, Spain.}
\address[Second]{Institute of Mathematical Sciences (CSIC-UAM-UCM-UC3M). Calle Nicol\'as Cabrera 13-15, Cantoblanco, 28049, Madrid, Spain.}
\begin{abstract}
Numerical methods that preserve geometric invariants of the system, such as energy, momentum or the symplectic form, are called geometric integrators. Variational integrators are an important class of geometric integrators. The general idea for those variational integrators is to discretize Hamilton’s principle rather than the equations of motion in a way that preserves some of the invariants of the original system. In this paper we  construct variational integrators with fixed time step for time-dependent Lagrangian systems modelling an important class of autonomous dissipative systems.  These integrators are derived via a family of discrete Lagrangian functions each one for a fixed time-step. This allows to recover at each step on the set of discrete sequences the preservation properties of variational integrators for autonomous Lagrangian systems, such as symplecticity or  backward error analysis for these systems. We also present a discrete Noether theorem for this class of systems. Applications of the results are shown for the problem of formation stabilization of multi-agent systems.

\end{abstract}


\end{frontmatter}

\section{Introduction}
Since the emergence of computational methods, fundamental properties such as accuracy, stability, convergence, and computational efficiency have been considered crucial for deciding the utility of a numerical algorithm. Geometric numerical integrators are concerned with numerical algorithms that preserve the system's fundamental physics by keeping the geometric properties of the dynamical system under study. The key idea of the structure-preserving approach is to treat the numerical method as a discrete dynamical system which approximates the continuous-time flow of the governing continuous-time differential equation, instead of focusing on the numerical approximation of a single trajectory. Such an approach allows a better understanding of the invariants and qualitative properties of the numerical method. Using ideas from differential geometry, structure-preserving integrators have produced a variety of numerical methods for simulating systems described by ordinary differential equations preserving its qualitative features. In particular, numerical methods based on discrete variational principles \cite{Lall1, mawest} may exhibit superior numerical stability and structure-preserving capabilities than traditional integration schemes for ordinary differential equations.

Variational integrators are geometric numerical methods derived from the discretization of variational principles \cite{mawest, hairer, Lall1}. These integrators retain some of the main geometric properties of the continuous systems, such as preservation of the manifold structure at each step of the algorithm, symplecticity, momentum conservation (as long as the symmetry survives the discretization procedure), and a good behavior of the energy function associated to the system for long time simulation steps. This class of numerical methods has been applied to a wide range of problems in optimal control \cite{objuma, leo,fer}, constrained systems \cite{Ley}, formation control of multi-agent systems \cite{CoGa1}, nonholonomic systems \cite{cortes}, accelerated optimization \cite{campos}, flocking control \cite{CoGa2} and motion planning for underactuated robots \cite{marin2}, among many others.

In this paper we construct variational integrators for non-autonomous Lagrangian systems with fixed time step (see \cite{mawest} for variable time step). More precisely, a variational integrator for a time-dependent Lagrangian system is derived through a family of discrete Lagrangian functions each one for a fixed time-step (see \cite{campos} and \cite{leo4}). This allows to recover at each step on the set of discrete sequences the preservation properties of variational integrators for autonomous Lagrangian systems such as symplecticity of the integrator or cosymplecticity of the modified time-dependent Hamiltonian system using  backward error analysis. We also obtain a discrete-time Noether Theorem for the relation between symmetries and first integrals. Such a result allow us to guarantee, for instance, an exponentially fast rate of change for the linear and angular momentum of certain mechanical systems. The class of variational integrators developed in this work are motivated by the recent applications of geometric integrators in contact \cite{contact1}, \cite{contact2}, celestial mechanics \cite{contact3} and formation control of multi-agent systems \cite{CoGa1, CoGa3}.

The remainder of the paper is structured as follows. Section $2$ introduces some geometric aspects of time-dependent Lagrangian systems, Noether symmetries, constants of the motion and its relation via a Noether Theorem for time-dependent Lagrangian systems. Section $3$ constructs the variational integrator for time-dependent Lagrangian systems and the discrete-time version of Noether theorem. In Section $4$ we derive the discrete Hamiltonian flow for discrete-time non-autonomous Hamiltonian Systems which is further employed in Section $5$ in the context of the backward error analysis.  Applications of the results are shown for the problem of formation stabilization of multi-agent systems are shown in Section $6$. Conclusions are presented in Section $7$.

\section{Symmetries and Constants of the Motion for Non-Autonomous Lagrangian Systems}\label{Sec2}
Let $Q$ be the configuration space of a mechanical system, that we will assume  is  a differentiable manifold of dimension $n$ with local coordinates $q=(q^1,\ldots,q^n)$. Let $TQ$ be the tangent bundle of $Q$, locally described by positions and velocities, $(q^i,\dot{q}^i)$ with $\hbox{dim}(TQ)=2n$. Let $T^{*}Q$ be its cotangent bundle, locally described by positions and momenta, $(q^{i},p_i)$ where also $\hbox{dim}(T^{*}Q)=2n$. The tangent and cotangent bundle at a point $q\in Q$ are denoted as $T_{q}Q$ and $T_q^*Q$, respectively. We denote by $\tau_Q: TQ\rightarrow Q$ the canonical projection on the tangent bundle which in local coordinates is given by $\tau_Q(q^i, \dot{q}^i)=(q^i)$ and by $\pi_Q: T^*Q\rightarrow Q$ the canonical projection on the cotangent bundle, 
$\pi_Q(q^i, p_i)=(q^i)$.

Consider a time-dependent Lagrangian $L:\R\times TQ\to\mathbb{R}$, and denote by $\F L\colon \R\times TQ\to \R\times T^*Q$ the Legendre transformation for $L$ given by 
 $(t,q,\dot q)\mapsto (t,q,p:=\partial L/\partial \dot q)$. We assume that $L$ is hyperregular, i.e. that $\F L$ is a diffeomorphism between $\R\times TQ$ and $\R\times T^*Q$. If $L$ is hyperregular, one can work out the velocities $\dot q=\dot q(t,q,p)$ in terms of $(t,q,p)$ and define the Hamiltonian function (the ``total energy'') $H\colon \R\times T^*Q\to \R$ as
$H(t,q,p)=p^T \dot q(t,q,p)- L(t, q,\dot q(t,q,p))$, where the inverse of the Legendre transformation to express $\dot q=\dot q(t,q,p)$ has been used. \vspace{.5em}

From the Lagrangian $L:\R\times TQ\to\mathbb{R}$ we can derive the Euler-Lagrange equations using  a variational principle, as follows. Denote by $C^2 (q_0,q_1)$ the set of twice differentiable curves with fixed end-points $q_0, q_1 \in Q$, that is, $C^2 (q_0,q_1)= \{q:[0,T]\longrightarrow Q| \ q \ \text{is} \ C^2, q(0)=q_0, q(T)=q_1\}$, and define the \textit{action functional} $\J:C^2 (q_0,q_1)\longrightarrow \R$, given by $q(\cdot)\mapsto \J(q(\cdot))=\int_{0}^{T} L(t, q(t),\dot{q}(t)) \ dt$. Critical points of this functional are described by the solutions of Euler-Lagrange equations, $\displaystyle{\frac{d}{dt}\left( \frac{\partial L}{\partial \dot{q}^i}\right) - \frac{\partial L}{\partial q^i}=0}$, that is, 

\begin{equation}\label{euler-Lagrange}
\frac{\partial^2 L}{\partial \dot{q}^i\partial \dot{q}^j}\ddot{q}^j
+\frac{\partial^2 L}{\partial \dot{q}^i\partial {q}^j}\dot{q}^j
+\frac{\partial^2 L}{\partial \dot{q}^i\partial t}- \frac{\partial L}{\partial q^i}=0.
\end{equation}
Since $L$ is  hyperregular, the matrix $\text{Hess}(L):=\left(\frac{\partial^2 L}{\partial \dot{q}^i \partial \dot{q}^j}\right)$ is non-singular. Hence, equations \eqref{euler-Lagrange} can be written as a system of explicit second-order time-dependent differential equations.

Two intrinsic geometrical objects (i.e., independent of the choice of local coordinates or the regularity of the Lagrangian), characterizing the tangent bundle $TQ$, are the Liouville vector field $\Delta$ and the vertical endomorphism $S$. These geometric objects allow, for instance, to describe the energy function of the system on the tangent bundle (instead of a Hamiltonian formalism on the cotangent bundle) and to describe the tangent bundle version of Noether theorem. Both can be regarded in a natural way as living on ${\mathbb R}\times TQ$ and we shall denote these extensions by the same symbols. In local coordinates, these geometrical objects can be written as $\Delta(q^i, v^i)=v^{i}\frac{\partial}{\partial \dot{q}^{i}}$ and $S(X^{i}\frac{\partial}{\partial q^{i}}+Y^i\frac{\partial}{\partial \dot{q}^{i}})=X^{i}\frac{\partial}{\partial \dot{q}^{i}}$.

By using the Liouville vector field we define the energy function $E_L$ on ${\mathbb R}\times TQ$ as $E_L=\Delta L-L$, or locally as $E_L=\dot{q}^i\frac{\partial L}{\partial \dot{q}^i}-L$. From equations (\ref{euler-Lagrange}) it follows that the energy, in general, is not preserved  in the non-autonomous case. In fact,  
\begin{equation}\label{energy}
\frac{d}{dt}E_L=-\frac{\partial L}{\partial t}\, .
\end{equation}

\begin{remark}
Alternatively, since $L$ is hyperregular, one can construct the energy function $E_L\colon \R\times TQ\to \R$ by using the Legendre transformation $\F L: TQ\rightarrow T^*Q$  \cite{Foundations} as  $E_L(t,q,\dot q)=\langle \F L(t,q,\dot q),\dot q\rangle - L(t,q,\dot q)$.
\end{remark}

Next, we define two lifts of vector fields on $Q$ to $TQ$.  
 Denote by ${\mathfrak X}(Q)$ the set of vector fields on $Q$ and let $X^{V}\in {\mathfrak X}(Q)$ the \textit{vertical lift} of $X\in {\mathfrak X}(Q)$, that is, the vector field on $TQ$ given by
\[
X^V(v_{q})=\left.\frac{d}{dt}\right|_{t=0}(v_{q}+tX(q))=(X(q))^V_{v_{q}},\; \ \forall v_{q}\in T_{q} Q.
\]
Locally,
$
X^V=X^i\frac{\partial}{\partial\dot{q}^i}
$
where $X=X^i\frac{\partial}{\partial q^i}$.

By denoting $\{\Phi^X_t\}$ the flow of a vector field $X\in {\mathfrak X}(Q)$, we can also define the \textit{complete lift}  $X^C\in {\mathfrak X}(TQ)$ of $X$  in terms of its flow which is the tangent lift  $\{T\Phi^X_t\}$. In other words, $X^C(v_q)=\left.\frac{d}{dt}\right|_{t=0}\left(T_q\Phi^X_t(v_q)\right)$. In coordinates,
$
X^C=X^i\frac{\partial}{\partial q^i}+\dot{q}^j\frac{\partial X^i}{\partial q^j}\frac{\partial}{\partial \dot{q}^i} \ .
$
As before, we denote by the same symbols  the corresponding extensions to ${\mathbb R}\times TQ$. Therefore, $X^V (t, v_q)=(0_t, X^V(v_q))$ and $X^C (t, v_q)=(0_t, X^C(v_q))$.

Using the vertical and complete lifts the Euler-Lagrange equations can be alternatively described  as follows \cite{mestdag-crampin, Diaz}. A curve $q(t)$ is a solution of Euler-Lagrange equations for $L$ if and only if 
\begin{equation}\label{qw}
\frac{d}{dt}\left(X^{V}(L)(q(t),\dot{q}(t))\right)=X^{C}(L) (q(t),\dot{q}(t)), \quad \forall \ X\in \mathfrak{X}(Q).
\end{equation}

In this paper we are only interested in symmetries  that come from vector fields on $Q$. This motivates the following definitions.

\begin{definition}\label{def:symmetry-lagrangian}
A vector field $X\in {\mathfrak X}(Q)$ is said to be a {\bf symmetry of the Lagrangian} $L: \R\times TQ\rightarrow \R$ if 
\[
X^C(L)=0.
\]
\end{definition}

Denoting by $d_{T}f: {\mathbb R}\times TQ\rightarrow {\mathbb R}$ the differential of a function $f:\mathbb{R}\times Q\to\mathbb{R}$, that is, $d_{T}f=\frac{\partial f}{\partial t}+\dot{q}^i\frac{\partial f}{\partial q^i}$ we can define  a more general class of symmetries called Noether symmetries.  

\begin{definition}
A vector field $X\in {\mathfrak X}(Q)$ is said to be a {\bf Noether symmetry} of  $L: \R\times TQ\rightarrow \R$ if 
\begin{equation}\label{lift}
X^C(L)=d_T f,
\end{equation}
for some function $f\in C^{\infty}(\R\times Q)$. 
\end{definition}
Observe that symmetries of the Lagrangian are a particular type of Noether symmetries with $f=0$ (or $f=\hbox{constant}$, in general).

From the Euler-Lagrange equations \eqref{qw}, together with \eqref{lift}, it follows the celebrated Noether theorem for the relation between symmetries and first integrals.

\begin{theorem}[\bf Noether Theorem] \label{noether1}
If $X$ is a Noether symmetry, that is   $X^C(L)=d_{T}f$. Then, $X^V(L)-f$ is a constant of the motion for the Euler-Lagrange equations for $L$. 
\end{theorem}

Next, assume that $G$ is a Lie group with Lie algebra ${\mathfrak g}$ and $\Phi: G\times Q\rightarrow Q$ a smooth left action of $G$ on $Q$. The infinitesimal generator $\xi_Q\in {\mathfrak X}(Q)$  corresponding to an element $\xi\in {\mathfrak g}$ is defined by (see, for instance, \cite{Bloch} Section $2.8$)
\begin{equation}\label{inf}
\xi_Q(q)=\frac{d}{ds}\Big|_{s=0}
\Phi(\hbox{exp}(s\xi), q).
\end{equation}
Denote by $\{\Phi^{\xi_Q}_s\}$ the flow of $\xi_Q$ then $\{T\Phi^{\xi_Q}_s\}$ is the flow of $\xi_Q^C$. The Lie group $G$ is said to be a \textbf{Lie group of symmetries} for $L$ if for all $\xi\in {\mathfrak g}$ and $s$, $L(t, T\Phi^{\xi_Q}_s(v_q))=L(t, v_q)$. Infinitesimally, the previous condition is equivalent to \begin{equation}\label{xicl}\xi_Q^C(L)=0\quad \hbox { for all } \quad \xi\in {\mathfrak g}  
\end{equation}
That is, if for any $\xi\in {\mathfrak g}$ we have that $\xi_Q$ is a symmetry of the Lagrangian as in Definition \ref{def:symmetry-lagrangian}.

As a consequence of Noether Theorem \ref{noether1} we deduce that for all $\xi \in {\mathfrak g}$ we have that such that $\xi_Q^V(L)$ is a constant of the motion for the Euler-Lagrange equations for $L$. 

\begin{example}\label{example}
	Consider the Lagrangian function $\mathbf{L}:T\mathbb{R}^{n}\equiv\mathbb{R}^{n}\times\mathbb{R}^{n}\to\mathbb{R}$ given by	\begin{equation}\label{Lagrangian}
	\mathbf{L}(q, \dot{q})=\frac{1}{2}||\dot{q}||^2-V(q), 
\end{equation} $q=(q_1,\ldots q_n)\in\mathbb{R}^{n}$ and $V:\mathbb{R}^{n}\to\mathbb{R}$ is a potential function which is assumed to be $SE(n)$-invariant. 

Next, consider the non-autonomous Lagrangian $L: {\mathbb R}\times TQ\rightarrow {\mathbb R}$ given by $L(t, q, \dot{q})=e^{-\kappa t} \mathbf{L}(q,\dot{q})$. The corresponding Euler-Lagrange equations for $L$ are
\begin{equation}
\ddot{q}_i=\kappa \dot{q}_i-\nabla_{q_i}{V}
                \label{doubleint}.
\end{equation}

In this case we have the energy of $L$ and ${\mathbf L}$ are related by 
$E_L=e^{-\kappa t}  E_{\mathbf L}$. 
Therefore using Equation (\ref{energy}) it follows that
$\displaystyle{
\frac{d E_{L}}{dt}={\kappa}{L}}$, indicating that the energy is not conserved along the evolution of the system. But, more intereting is to observe that 
$\displaystyle{
\frac{d E_{\mathbf L}}{dt}={\kappa}||\dot{q}||^2}$
and therefore we have dissipation of energy if $k<0$, preservation if $k=0$ and energy growth if $k>0$.

 The time-dependent Lagrangian $L$ is $SE(n)$-invariant under the Lie group action $\Phi: SE(n)\times {\mathbb R}^{n}\rightarrow {\mathbb R}^{n}$
	given by $\Phi(R, q)=\Phi(R, q_1, \ldots, q_n)=(Rq_1, Rq_2, \ldots, Rq_n)$ where $q_a\in {\mathbb R}$, $1\leq a\leq n$. That is, 
	\[
	L(t, Rq_1,  \ldots, Rq_n, R\dot{q}_1,  \ldots, R\dot{q}_n)=
		L(t, q_1, \ldots, q_n, \dot{q}_1,  \ldots, \dot{q}_n).
	\]
	Infinitesimally this invariance  means that $\xi_Q^C(L)=0$ for any $\xi\in SE(d)$. Using that $\xi_Q^C(q)(L)=e^{-{\kappa}t} \xi_Q^C(q)(\mathbf{L})$, then 
	$\xi_Q^C(\mathbf{L})=0$. Therefore, 
	by Noether Theorem \ref{noether1} it follows that  $\xi^V_Q(L)=(e^{-{\kappa}t} \xi^V_Q(q)(\mathbf{L}))$ are constants of the motion  for all $\xi\in {\mathfrak g}$ for the system given by equations \eqref{doubleint}.
	 As a consequence, if $k<0$, we deduce  the exponential decay of the functions $J_{\xi}:=\xi^V_Q(q)(\mathbf{L})$:
	\begin{equation}\label{decay}
	||J_{\xi}(q(t), \dot{q}(t))||=e^{-\kappa t}	||J_{\xi}(q(0), \dot{q}(0))||.
	\end{equation} Note that in the case $d=3$, we have two types of infinitesimal generators:

	     \textbf{[(a)]}  Translation in the direction ${\mathbf a}\in {\mathbb R}^n$ makes the Lagrangian $SE(n)$-invariant. In this case, the infinitesimal generator is given by $\displaystyle{\xi_Q= \displaystyle{{\mathbf a}\cdot \frac{\partial}{\partial q}}}$. Therefore, by \eqref{decay} the \textit{linear momentum} $\displaystyle{J_
	     {\xi}=\xi^V_Q(q)(\mathbf{L})= {\mathbf a}\cdot \dot{q}}$ decays exponentially.
	     
	     \textbf{[(b)]} Rotations in the system about some fixed axis makes the Lagrangian $L$ also $SE(n)$-invariant. For instance, with $n=3$, by considering rotations along the $z$-axis, the infinitesimal generator is given by the vector field
	     $\displaystyle{\xi_Q=\left(x\frac{\partial}{\partial y}- y\frac{\partial}{\partial x}\right)}$.
	     In this case, by \eqref{decay}, the quantity which exponentially decays is the \textit{angular momentum} $\displaystyle{J_{\xi}=\xi^V_Q(q)(\mathbf{L})= x\dot{y}-y\dot{x}}$.

\end{example}

\section{Symmetries and Constants of the Motion for Discrete-Time Non-Autonomous Mechanical Systems}

Variational integrators (see \cite{mawest} for details) are derived
from a discrete variational principle. These integrators  retain
some of the main geometric properties of the continuous-time systems,
such as symplecticity, momentum conservation (as long as the
symmetry survives the discretization procedure), and good (bounded) behavior
of the energy associated to the system (see \cite{hairer} and
references therein). 

A \textit{discrete Lagrangian} is a differentiable function
$L_d\colon Q \times Q\to \R$, which may be considered as an
approximation of the action integral defined by a continuous regular 
Lagrangian $L\colon TQ\to \R.$ That is, given a time step $h>0$
small enough, $\displaystyle{L_d(q_0, q_1)\approx \int^h_0 L(q(t), \dot{q}(t))\; dt}$, where $q(t)$ is the unique solution of the Euler-Lagrange equations
for $L$ with  boundary conditions $q(0)=q_0$ and $q(h)=q_1$.

Construct the grid $\{t_{k}=kh\mid k=0,\ldots,N\},$ with $Nh=T$
and define the discrete path space
$\mathcal{C}_{d}:=\{q_{d}:\{t_{k}\}_{k=0}^{N}\ra Q\}.$ We
identify a discrete trajectory $q_{d}\in\mathcal{C}_{d}$ with its
image $q_{d}=\{q_{k}\}_{k=0}^{N}$, where $q_{k}:=q_{d}(t_{k})$. Define 	$$
	\mathcal{C}_d(q_0,q_N)=\left\{ q_d:\left\{ k \right\}_{k=0}^N \to Q\mid q_d(0)=q_0, q_d(N)=q_N\right\}.
	$$ The
discrete action $\mathcal{A}_{d}:{C}_d(q_0,q_N)\ra\R$ for a
sequence $q_d$ is calculated by summing the discrete Lagrangian on each
adjacent pair and is defined by \begin{equation}\label{acciondiscreta}
\mathcal{A}_d(q_{d}) = \mathcal{A}_d(q_0,...,q_N) :=\sum_{k=0}^{N-1}L_d(q_k,q_{k+1}).
\end{equation}
For any product manifold $Q_1\times Q_2,$
$T^{*}_{(q_1,q_2)}(Q_1\times Q_2)\simeq T^{*}_{q_1}Q_1\oplus
T^{*}_{q_2}Q_2,$ for $q_1\in Q_1$ and $q_2\in Q_2$ where $T^{*}Q$
denotes the cotangent bundle of a differentiable manifold $Q.$
Therefore, any covector $\alpha\in T^{*}_{(q_1,q_2)}(Q_1\times Q_2)$
admits an unique decomposition $\alpha=\alpha_1+\alpha_2$ where
$\alpha_i\in T^{*}_{q_i}Q_i,$ for $i=1,2.$ Thus, given a discrete
Lagrangian $L_d$ we have the following decomposition
  $dL_{d}(q_0,q_1)=D_{1}L_d(q_0,q_1)+D_{2}L_d(q_0,q_1)$, where $D_{1}L_d(q_0,q_1)\in T^*_{q_0}Q$ and $D_{2}L_d(q_0,q_1)\in
T^*_{q_1}Q$. Discrete Euler Lagrange equations (see \cite{mawest} for instance) are given by a critical sequence for $\mathcal{A}_d$ on the space $\mathcal{C}_d(q_0,q_N)$.  That is, the {\it discrete Euler-Lagrange equations} are $$D_1 L_d(q_{k+1},q_{k+2})+D_2L_d(q_{k},q_{k+1})=0,\,\forall k=0,\ldots,N-2,$$ where $D_1$ and $D_2$ denote the partial derivatives with respect to the first and second component of $L_d$, respectively.

For non-autonomous systems \cite{campos, leo4} we introduce, in the discrete setting,  a family of maps
	$L^k_d: Q\times Q\rightarrow {\mathbb R}$, $k=0,\ldots, N-1$ where we are now fixing the number of steps $N\in\mathbb{N}$ and considering a discrete Lagrangian on the set of discrete sequences defined on each step $q_d:\{k\}_{k=0}^{N}\to Q$.

	The family of discrete Lagrangians $\{L^k_d\}_{k=0}^{N-1}$ will be called {\it discrete time-dependent Lagrangian} and simply denoted by $L_d^k$.
	
We look for the extremals of the corresponding  discrete action given  by
	$
	\displaystyle{S_d (q_d)=\sum_{k=0}^{N-1}L^k_d(q_k,q_{k+1}).
	}$
	The stationary condition for variations vanishing at the end points of the discrete sequences gives rise to the {\it discrete Euler-Lagrange equations} \cite{leo4}
	\begin{equation}\label{dfeleq}
	D_1 L^{k+1}_d (q_{k+1},q_{k+2})+D_2L^{k}_d(q_{k},q_{k+1})=0,\, k=0,\ldots,N-2. 
	\end{equation}

  The discrete Euler-Lagrange equations implicitly defines a family of local discrete flows $\{\Psi_d^{k,k+1}\}_{k=0}^{N-2}$ as
\begin{equation}\label{discrete-flow}
\begin{array}{cccc}
     \Psi_d^{k,k+1}:&  Q\times Q&\longrightarrow & Q\times Q \\
     &  (q_k, q_{k+1})&\longmapsto &(q_{k+1}, q_{k+2}(q_k,q_{k+1},k))
\end{array}
\end{equation}
where  $q_{k+2}$ is locally well defined by using the discrete Euler-Lagrange equations and assuming the non-singularity of the matrix $D_{12}L_d^k(q_k, q_{k+1})$ for each $k$ and $(q_k, q_{k+1})\in Q\times Q$.  Observe that the map $\Psi_d^{k,k+1}$ transforms a point  $(q_k, q_{k+1})$ at a discrete time $k$ to a new point $(q_{k+1}, q_{k+2})$ now at discrete time $k+1$.

Equations \eqref{dfeleq} define the integration scheme $(q_{k-1},q_k)\mapsto(q_k,q_{k+1}).$ By defining the discrete (post and pre) momenta
  \begin{align}
  p^{+}_{k}:=&D_2L_d^{k-1}(q_{k-1},q_k),\, k=1,\ldots,N\label{momentum1}\\
  p^{-}_{k}:=&-D_1L_d^k(q_{k},q_{k+1}),\,k=0,\ldots,N-1,\nonumber
  \end{align} equations \eqref{dfeleq} lead to the integration scheme $(q_k,p_k)\mapsto(q_{k+1},p_{k+1})$, by writing \eqref{dfeleq} as  $p_{k}^{-}=p_{k}^{+}$.

Given a vector field $X\in {\mathfrak X}(Q)$ we can define the vector fields $X^{C, d}$ and $X^{V, d}$ in $X\in {\mathfrak X}(Q\times Q)$ by $X^{C,d}(q_0, q_1)=(X(q_0), X(q_1))$ and $X^{V,d}(q_0, q_1)=(X(q_0), 0_{q_1})$. In terms of these vector fields, the discrete Euler-Lagrange equations can be writen similarly to (\ref{qw}), as (see \cite{Diaz} for details) 
\begin{equation}\label{qert}
X^{C,d}(q_k, q_{k+1})(L^k_d)=\left(X^{V,d}(q_{k}, q_{k+1})(L^k_d)-X^{V,d}(q_{k+1}, q_{k+2})(L^{k+1}_d)\right),
\end{equation}$\forall X\in {\mathfrak X}(Q),\,k=0,\ldots,N-2$. 

\begin{definition}\label{def:symmetry-lagrangian-d}
A vector field $X\in {\mathfrak X}(Q)$ is said to be a {\bf  symmetry of the discrete time-dependent Lagrangian} $L^k_d: Q\times Q\rightarrow \R$ if for each $k\in\{0,\ldots,N-1\}$,
\[
X^{C,d}(L^k_d)=0.
\]
\end{definition}

For a family of functions $f^k: Q\rightarrow \R$, $k\in\{0,\ldots,N-1\}$
define $d^k_T f: Q\times Q\rightarrow \R$ by 
\[
d^k_T f (q_k, q_{k+1})=f^{k+1}(q_{k+1})-f^k(q_k).
\] Then, we can define Noether symmetries for the discrete-time Lagrangian $L^k_d$ as follows.

\begin{definition}
A vector field $X\in {\mathfrak X}(Q)$ is said to be a {\bf discrete Noether symmetry} of  $L^k_d: Q\times Q\rightarrow \R$ if 
\begin{equation}\label{completdisc}
X^{C,d}(L^k_d)=d^k_T f
\end{equation}
for each $k\in\{0,\ldots,N-1\}$ and for a family of functions $f^k: Q\rightarrow \R$. 
\end{definition}

In the same way as the continuous-time case, as a  consequence of the discrete Euler-Lagrange equations \eqref{qert}, together with \eqref{completdisc}, we deduce Noether Theorem for the relation between symmetries of the discrete Lagrangian and first integrals of the discrete Euler-Lagrange equations.

\begin{theorem}[\bf Discrete Noether Theorem] \label{noether1-d}
If $X\in\mathfrak{X}(Q)$ is a discrete Noether symmetry for the discrete-time Lagrangian $L^k_d$, that is  $X^{C,d}(L^k_d)=d^k_{T}f$, then,  $X^{V,d}(L^k_d)-f^k$ is a constant of the motion for the discrete Euler-Lagrange equations for $L^k_d$ for each $k$, $k=0,\ldots,N-1$. 
\end{theorem}

As in Section \ref{Sec2}, consider the action of a Lie group $G$ on $Q$, $\Phi: G\times Q\rightarrow Q$, with infinitesimal generator as \eqref{inf}. This action can be
lifted to $Q\times Q$ by $\Phi^{Q\times Q}_{g}(q_0,q_1)=(\Phi_{g}(q_0), \Phi_g(q_1))$ which has
an infinitesimal generator $\xi_{Q\times Q}:Q\times Q\to T(Q\times Q)$ given by $\xi_{Q\times Q}(q_0,q_1)=(\xi_Q(q_0),\xi_Q(q_1))=\xi_Q^{C,d}(q_0, q_1)$.

Assume that the family of discrete Lagrangians $L^k_d$ is invariant under the lifted action, that is, for all $g\in G$
\[
L^k_d\circ\Phi^{Q\times Q}_{g}(q_0,q_1)=L^k_d(q_0, q_1), \forall (q_0, q_1)\in Q\times Q\; .
\]
Infinitesimally, this is equivalent to 
\begin{equation}\label{infdisccom}
(\xi_{Q\times Q})^{C, d}(L_d^k)=0,\,\hbox{for all }\xi \in {\mathfrak g}.\end{equation} 
That is $\xi_Q$ is symmetry of the discrete Lagrangian $L_d^k$.

From Equation (\ref{qert}) and \eqref{infdisccom} we obtain a discrete-time version of Noether Theorem as follows
\begin{theorem} \label{noetherd}
	Let $G$ be a Lie group of symmetries for $L^k_d$, that is, $(\xi_{Q\times Q})^{C,d}(L^k_d)=0$ for all $k$ and $\xi \in {\mathfrak g}$. Then, $(\xi_{Q\times Q})^{V, d}(L^k_d)$ is a constant of the motion for the discrete Euler-Lagrange equations for $L^k_d$.
\end{theorem}

\begin{example}\label{example2}

Consider the time-dependent Lagrangian function $L:\mathbb{R}\times\mathbb{R}^{n}\times\mathbb{R}^{n}\to\mathbb{R}$ given in Example \ref{example} by  \begin{equation}\label{stabletimedependet}L(t,q,\dot{q})=e^{-\kappa t}\left(\frac{1}{2}||\dot{q}
||^{2}-V(q)\right).\end{equation}

To construct the geometric integrator, the velocities are discretized by finite-differences, i.e., $\displaystyle{\dot{q}_i=\frac{q_{k+1}^i-q_k^i}{h}}$ for $t\in[t_k,t_{k+1}]$. The discrete Lagrangian $L_{d,h}^{k}:\mathbb{R}^{n}\times\mathbb{R}^{n}\to\mathbb{R}$ is given by setting the trapezoidal discretization for the time-dependent Lagrangian $L$ given by \eqref{stabletimedependet}, that is, \begin{align*}
L_{d,h}^{k}(q_k,q_{k+1})=\frac{h}{2}L\left(kh, q_k,\frac{q_{k+1}-q_k}{h}\right)+\frac{h}{2}L\left((k+1)h,q_{k+1},\frac{q_{k+1}-q_k}{h}\right)\end{align*}where, $h>0$ is the time step. 

The discrete Euler-Lagrange equations for $L_{d,h}^{k}$ are given by \begin{align}
    0=&(q_{k+1}-q_k)e^{-\kappa(kh)}-(q_{k+2}-q_{k+1})e^{-\kappa h(k+2)}\label{deleq}\\
&-e^{-\kappa h(k+1)}(q_k-2q_{k+1}+q_{k+2}+h\nabla_{q_{k+1}}V(q_{k+1}))\nonumber.
\end{align}

After some calculus we can write equations \eqref{deleq} as the following explicit integration scheme \begin{equation}\label{integrator-explicit}
    q_{k+2}=\hat{\kappa}_{h}q_{k+1}-\kappa_{h}q_k-h\bar{\kappa}_{h}\nabla_{q_{k+1}}V(q_{k+1}),
\end{equation}with $\displaystyle{\kappa_{h}=\frac{e^{\kappa hk}+1}{e^{-\kappa hk}+1}}$, $\displaystyle{\bar{\kappa}_{h}=\frac{1}{e^{-\kappa hk}+1}}$, $\displaystyle{\hat{\kappa}_{h}=\frac{2+e^{-\kappa hk}+e^{\kappa hk}}{e^{-\kappa hk}+1}}$.

Note that the previous equations are a set of $n(N-1)$ for the $n(N+1)$ unknowns $\{q_k\}_{k=0}^{N}$. Nevertheless the boundary conditions on initial positions and velocities  $q_0=q(0)$, $v_{q_0}=\dot{q}(0)$ contribute to $2n$ extra equations that convert eqs. \eqref{deleq} into a nonlinear root finding problem of $n(N-1)$ equations and the same amount of unknowns. To start the algorithm we use the boundary conditions for the first two steps, that is, $q_0=q(0)$ and $q_1=hv_{q_0}+q_{0}=h\dot{q}(0)+q(0).$

The energy function is also discretized by using a trapezoidal discretization.  In particular, the energy $E_L:\mathbb{R}\times TQ\to \mathbb{R}$ is given by $$E_L(t,q,\dot{q})=e^{-\kappa t}\left(\frac{1}{2}||\dot{q}||^2+V(q)\right).$$ Using the trapezoidal rule for $E_L$, the discrete energy function  $E_d:\mathbb{R}^n\times\mathbb{R}^n\to\mathbb{R}$ is given by \begin{equation}E_d(q_k,q_{k+1})=\frac{1}{2h}||q_{k+1}-q_k||^2(e^{-\kappa kh}+e^{-\kappa(k+1)h})+\frac{h}{2}(e^{-\kappa hk}V(q_k)+e^{-\kappa(k+1)h}V(q_{k+1})).\end{equation}

Next we show the performance of the proposed variational integrator in numerical simulations. For simplicity we consider $Q=\mathbb{R}^3$ and $V(q)=0$. Initial positions were arbitrary selected as $q_{0}=[18,6,10]$ and we set the initial velocities to be $v_{0}= [2.22 , -1.86,  3.48]$. Note that by using the fact that $\dot{E}_{L}=\kappa L=\kappa E_L$, for $\kappa<0$ the energy of the system decays exponentially, so, for simulation results we choose as damping gain $\kappa=-5$. The simulation for the enegy behaviour was conducted with an end time of $1$ seconds and time steps of $h=0.005$ seconds, which results in $N=200$ iterations. In Figure \ref{fig:enerna}, we show the exponential decay for the rate of change of the total energy function of the system, in both case, for the non-autonomous energy function (left figure) and the autonomous energy function (right figure).

\begin{figure}[h!]
    \centering
    \includegraphics[scale=0.5]{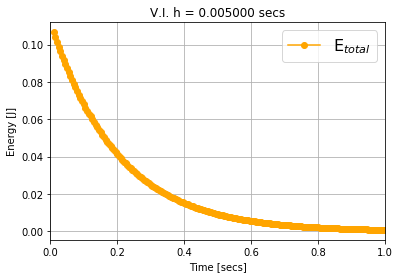}    \includegraphics[scale=0.5]{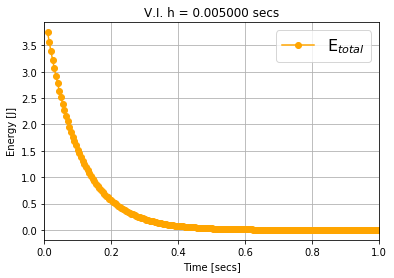}
    \caption{Exponential decay for the rate of change of the total energy of system. The left figure shows the evolution of the non-autonomous energy function while the right figure corresponds with the evolution of the autonomous energy function.}
    \label{fig:enerna}
\end{figure}

Observe also that the Lagrangian $L_{d,h}^{k}$ is $SE(d)$-invariant, therefore applying the discrete Noether Theorem \ref{noetherd}, it follows that for all 
$\xi \in \mathfrak{se}(d)$,  
\begin{equation}\label{momentumvertical}\xi^{V, d}_Q (q_k, q_{k+1})(L_{d,h}^{k})=\xi^{V, d}_Q (q_{k+1}, q_{k+2})(L_{d,h}^{k+1}),\end{equation} for all $k=0, \ldots, N-1$ and where $\{q_k\}$ is a solution of the discrete Euler-Lagrange equations.  

\begin{figure}[h!]
\centering
\includegraphics[width=0.44\columnwidth]{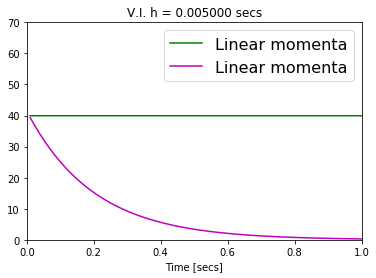}
\includegraphics[width=0.45\columnwidth]{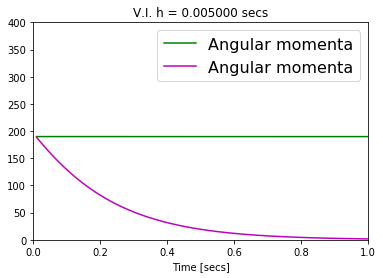}
	\caption{Exponential decay of the constants of the motion and preservation of the linear (left) and angular momentum (right). The green lines show the preservation of these  quantities in the non-autonomous case while the purple curves show the decay in the autonomous situation.}
\label{fig: momentum}
\end{figure}

Figure \ref{fig: momentum} shows an application of Noether Theorem \ref{noetherd}. In particular, Figure \ref{fig: momentum} (left figure) shows the preservation of the associated ``linear momentum" given by $$-D_{1}L_{d,h}^{k}(q_k,q_{k+1})=e^{-\kappa kh}\left((1+e^{-kh})\left(\frac{q_{k+1}-q_k}{h}\right)-\frac{h}{2}\nabla_{q_k}V(q_k)\right),$$ and the exponential decay of the constants of the motion. Similar simulation results can obtained for angular momentum as shown in Figure \ref{fig: momentum} (right figure). Note that in the case of the associated ``angular momentum", it is given by $$e^{-\kappa kh}\left((1+e^{-kh})\left(\frac{q_{k+1}-q_k}{h}\right)-\frac{h}{2}\nabla_{q_k}V(q_k)\right)\times q_{k+1}.$$

\end{example}

\section{Discrete Hamiltonian Flow for Discrete-Time Non-Autonomous Mechanical Systems}\label{sec: ham}

Consider $L:\R\times TQ\to\mathbb{R}$ as in Section \ref{Sec2}. Since $L$ is hyperregular we can determine the Hamiltonian function $H:\R\times T^{*}Q\to\mathbb{R}$ by using the Legendre transform $\mathbb{F}L:\R\times TQ\to\mathbb{R}\times T^{*}Q$ by
\[
  H=E_L\circ (\mathbb{F}L)^{-1}=p^T \dot q(t,q,p)- L(t, q,\dot q(t,q,p)) \,,
\]
which induces the cosymplectic structure $(\eta, \Omega_H)$ on $T^*Q\times\mathbb{R}$ with $\Omega_H=-d(\text{pr}_1^*\theta_Q-H \eta)=\Omega_Q+d H\wedge\dt$ and $\eta=\text{pr}_2^*\dt\;$, where \(\text{pr}_i\), \(i=1,2\), are the projections to each  factor and \(\theta_Q\) denotes the Liouville 1-form on \(T^*Q\) \cite{Foundations}, given in induced coordinates by \(\theta_Q=p_i\, dq^i\). We also denote by \(\Omega_Q=-d\text{pr}_1^*\theta_Q\) the pullback of the canonical symplectic 2-form \(\omega_Q=-d\theta_Q\) on \(T^*Q\). In coordinates, \(\Omega_Q=dq^i\wedge d p_i\) but observe that now \(\Omega_Q\) is presymplectic since \(\ker \Omega_Q=\text{span}\{\partial/\partial t\}\).
Therefore in induced coordinates \((t, q^i, p_i)\):
\[
  \Omega_H=dq^i\wedge d p_i+d H\wedge\dt\,,\qquad \eta=\dt.
\]

We define the \textbf{evolution vector field} \(E_H\in \mathfrak{X}(T^*Q\times \R)\) by
\begin{equation}
  \label{eq:qqq}
	i_{E_H}\Omega_H=0\; ,\qquad i_{E_H}\eta=1
\end{equation}
In local coordinates the evolution vector field is:
\[
  E_H=\frac{\partial}{\partial t}+\frac{\partial H}{\partial p_i}\frac{\partial}{\partial q^i}-\frac{\partial H}{\partial q^i}\frac{\partial}{\partial p_i}.
\]
The integral curves of \(E_H\) are given by:
\begin{equation}\label{tdepham}
  \dot t=1\,,\qquad \dot q^i=\frac{\partial H}{\partial p_i} \,, \qquad \dot p_i=-\frac{\partial H}{\partial q^i}\,.
\end{equation}

From Equation \eqref{eq:qqq} we deduce that the flow of \(E_H\) verifies the preservation relations 
\begin{equation}
  \label{eq:kop}
  \mathcal{L}_{E_H}( \Omega_Q+dH\wedge\dt)=0,\, \qquad \mathcal{L}_{E_H}\eta=0.
\end{equation}

The integral curves of \(E_H\) are precisely the curves of the form \(t\mapsto \mathbb{F}L(\sigma'(t),t)\) where \(\sigma\colon I\to Q\) is a solution of the Euler-Lagrange equations for the time-dependent Lagrangian \(L\colon \R\times TQ\to \R\).
 
 Denote by \(\Psi_s\colon \mathcal{U}\subset T^* Q\times \R\to T^* Q\times \R\) the flow of the evolution vector field \(E_H\), where \(\mathcal{U}\) is an open subset of \(T^*Q\times \R\).
Observe that $\Psi_s ( \alpha_q, t)=(\Psi_{t,s}(\alpha_q), t+s), \, \alpha_q\in T_q^* Q$, where \(\Psi_{t,s}(\alpha_q)=\text{pr}_1 (\Psi_s ( \alpha_q, t))\).
Therefore from the flow of \(E_H\) we induce a map
\[
\Psi_{t,s}\colon \mathcal{U}_t\subseteq T^*Q\to T^*Q
\]
where \(\mathcal{U}_t=\{ \alpha_q\in T^*Q \;|\; (\alpha_q, t)\in \mathcal{U}\}\). Observe that if we know \(\Psi_{t,s}\) for all \(t\), we can recover the flow \(\Psi_s\) of \(E_H\).

From equations \eqref{eq:kop} we have that $\Psi_s^*(\Omega_Q+ dH\wedge\dt)=\Omega_Q+dH\wedge\dt$ and $\Psi_s^*(\eta)=\eta\;$. 
The previous preservation properties are associated with the symplecticity of the family of maps \(\{\Psi_{t,s}\colon T^*Q\to T^*Q\}\). In particular, for all $t,s$ with $s$ small enough it has been show in \cite{campos} that $\Psi_{t,s}:\mathcal{U}_t\subseteq T^*Q\to T^*Q$ is a symplectomorphism, that is, $\Psi_{t,s}^*\omega_Q=\omega_Q$.

Given a discrete Lagrangian $L_d^k:Q\times Q\to\mathbb{R}$, the \textit{discrete Legendre transformations} $\mathbb{F}_{L_d^k}^{\pm}:Q\times Q\to T^{*}Q$ are defined at each $k$ through the momentum equations \eqref{momentum1} as  \begin{small}\begin{align}
\mathbb{F}_{L_d^k}^{+}(q_0,q_1)=&(q_1,D_2L_d^k(q_0,q_1))=(q_1,p_1)\label{fld1}\\
\mathbb{F}_{L_d^k}^{{-}}(q_0,q_1)=&(q_0,-D_1L_d^{k}(q_0,q_1))=(q_0,p_0).\label{fld2}
\end{align}\end{small}

If for each $k$ both discrete Legendre transformations are locally diffeomorphisms for
nearby $q_0$ and $q_1$, then we say that $L_ d^k$ is
\textit{regular}.  Using $\mathbb{F}_{L_d^k}^{{\pm}}$, the discrete Euler--Lagrange
equations \eqref{dfeleq} can be written as
$$\displaystyle{\mathbb{F}_{L_d^{k+1}}^{{-}}(q_{k+1},q_{k+2})=\mathbb{F}_{L_d^k}^{+}(q_{k},q_{k+1})}.$$

Consider $\Psi_d^{k,k+1}\colon Q\times Q \to Q\times Q$ defined by \eqref{discrete-flow}. It will be useful to note that

\begin{equation}\label{relationF}\mathbb{F}_{L_d^k}^{+}=\mathbb{F}_{L_d^{k+1}}^{-}\circ \Psi_d^{k,k+1}.\end{equation}

\begin{definition}\label{defhamin}
 We define the \textbf{discrete Hamiltonian flow} $ \widetilde{\Psi}_d^{k,k+1}:T^{*}Q\to T^{*}Q$ as \begin{equation}\label{sympint}
\widetilde{\Psi}_d^{k,k+1}=\mathbb{F}_{L_d^{k+1}}^{-}\circ\Psi_d^{k,k+1}\circ\left(\mathbb{F}_{L_d^{k+1}}^{-}\right)^{-1},\quad \widetilde{\Psi}_d^{k,k+1}(q_0,p_{0})=(q_1,p_{1}).
\end{equation} \end{definition}

Alternatively, it can also be defined as \begin{equation}\label{sympint+}
\widetilde{\Psi}_d^{k,k+1}=\mathbb{F}_{L_d^k}^{+}\circ\Psi_d^{k,k+1}\circ\left(\mathbb{F}_{L_d^k}^{+}\right)^{-1},\quad \widetilde{\Psi}_d^{k,k+1}(q_0,p_{0})=(q_1,p_{1}).\end{equation}

In analogy with \cite{mawest} we have the following results:

\begin{proposition}\label{Theo2}
The diagram in Figure \ref{fig:discretemaps} is commutative.
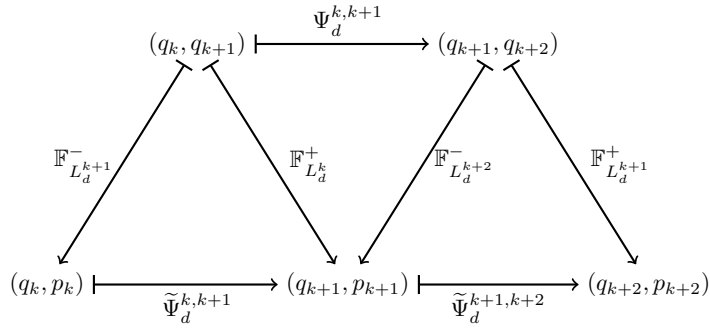
\begin{figure}[htbp]
\begin{center}
\begin{tikzpicture}[scale=0.8, every node/.style={scale=0.9}]

\path (2.5,4) node(a) {$(q_k,q_{k+1})$}; \path
(7.5,4) node(b) {$(q_{k+1},q_{k+2})$}; \path (0,0)
node(c) {$(q_k,p_k)$}; \path (5,0) node(d)
{$(q_{k+1},p_{k+1})$}; \path (10,0) node(e)
{$(q_{k+2},p_{k+2})$};

\path (5.0,4.0) node[anchor=south] (f) {$\Psi_d^{k,k+1}$}; \path (1.25,2)
node[anchor=east] (g) {$\mathbb{F}_{L_d^{k+1}}^{-}$}; \path (3.9,2)
node[anchor=west] (h) {$\mathbb{F}_{L_d^{k}}^{+}$}; \path (6.3,2)
node[anchor=west] (i) {$\mathbb{F}_{L_d^{k+2}}^{-}$}; \path (8.9,2)
node[anchor=west] (j) {$\mathbb{F}_{L_d^{k+1}}^{+}$}; \path (2.5,0)
node[anchor=north] (k) {$\widetilde{\Psi}_d^{k,k+1}$}; \path (7.5,0)
node[anchor=north] (l) {$\widetilde{\Psi}_d^{k+1,k+2}$}; \draw[thick,black,|->]
(a) -- (b); \draw[thick,black,|->] (a)  -- (c);
\draw[thick,black,|->] (a)  -- (d); \draw[thick,black,|->] (b) --
(d); \draw[thick,black,|->] (b)  -- (e); \draw[thick,black,|->] (c)
-- (d); \draw[thick,black,|->] (d)  -- (e);

\end{tikzpicture}
\caption[Correspondence between the discrete Lagrangian and the
 discrete Hamiltonian map]{Correspondence between the discrete
Lagrangian and the discrete Hamiltonian flows.}\label{diagram:LH}
\label{fig:discretemaps}
\end{center}
\end{figure}

\end{proposition}
\textit{Proof Proposition \ref{Theo2}:}
The central triangle is \eqref{relationF}.
The
parallelogram on the left-hand side is commutative by \eqref{sympint}, so the triangle on the left is commutative. The triangle on the right is the same as the triangle on the left, with shifted indices. Then parallelogram on the right-hand side is commutative and therefore the triangle on the right-hand side.\hfill$\diamond$

\begin{corollary}\label{corollaryrelations} The following definitions of the discrete
Hamiltonian flow are equivalent:
$
\widetilde{\Psi}_d^{k,k+1}=\mathbb{F}_{L_d^k}^{+}\circ \Psi_d^{k,k+1}\circ(\mathbb{F}_{L_d^k}^{+})^{-1}$,\,
$\widetilde{\Psi}_d^{k,k+1}=\mathbb{F}_{L_d^{k+1}}^{-}\circ \Psi_d^{k,k+1}\circ(\mathbb{F}_{L_d^{k+1}}^{-})^{-1}$,\,
$\widetilde{\Psi}_d^{k,k+1}=\mathbb{F}_{L_d^k}^{+}\circ(\mathbb{F}_{L_d^{k+1}}^{-})^{-1}$.\end{corollary}

In addition, for each $k$ we have that $(\mathbb{F}^{+}L^{k}_{d})^{*}\omega_Q=(\mathbb{F}^{-}L^{k}_{d})^{*}\omega_Q \,$ (see \cite{mawest} and \cite{campos}), so, for each $k$, the discrete Hamiltonian flow $\widetilde{\Psi}_d^{k,k+1}$ is a \textbf{symplectic transformation}, that is $(\widetilde{\Psi}_d^{k,k+1})^*\omega_Q=\omega_Q\;$. Moreover, given the map $\widetilde{\Psi}_d^{k,k+1}(q_k, p_k)=(q_{k+1}, p_{k+1})$, we have the map $(kh,q_k, p_k)=((k+1)h, q_{k+1}, p_{k+1})$ on $\R\times T^*Q$ giving explicitly the information of the evolution of discrete time.

\begin{example}
    
    Continuating with Examples \ref{example} and \ref{example2}, by using that \begin{align*}
D_1L_{d,h}^k(q_k,q_{k+1})&=-\frac{q_{k+1}-q_{k}}{h}\left(e^{-\kappa kh}+e^{-\kappa(k+1)h}\right)-\frac{h}{2}e^{-\kappa kh}\nabla_{q_k}V(q_k),\\
D_2L_{d,h}^{k}(q_{k},q_{k+1})&=\frac{q_{k+1}-q_{k}}{h}\left(e^{-\kappa kh}+e^{-\kappa(k+1)h}\right)-\frac{h}{2}e^{-\kappa (k+1)h}\nabla_{q_{k+1}}V(q_{k+1}),
\end{align*} we define the  Legendre transformations as \begin{align*}
\mathbb{F}_{L_d^k}^{+}(q_k,q_{k+1})&=\left(q_{k+1},\frac{q_{k+1}-q_{k}}{h}\left(e^{-\kappa kh}+e^{-\kappa(k+1)h}\right)-\frac{h}{2}e^{-\kappa (k+1)h}\nabla_{q_{k+1}}V(q_{k+1})\right),\,\\
\mathbb{F}_{L_d^k}^{-}(q_k,q_{k+1})&=\left(q_k,\frac{q_{k+1}-q_{k}}{h}\left(e^{-\kappa kh}+e^{-\kappa(k+1)h}\right)+\frac{h}{2}e^{-\kappa kh}\nabla_{q_{k}}V(q_{k})\right). 
\end{align*}
Using the last two expressions and  $\Psi_d^{k,k+1}$ given by \eqref{integrator-explicit}, it follows the construction of the Hamiltonian flow $\widetilde{\Psi}_d^{k,k+1}$ by Corollary \eqref{corollaryrelations}.
\end{example}

\section{Backward Error Analysis for Discrete-time Non-autonomous Mechanical Systems}\label{sec: energy}

Next we will show the discrete Hamiltonian flow $\widetilde{\Psi}_d^{k,k+1}$ defined in \eqref{sympint} has an asymptotically correct decay behavior by studying the rate of decay of a truncated modified Hamiltonian function following the approach of \textit{Backward Error Analysis} \cite{hairer} (Chapter IX), \cite{hansen} (Sec. $4$)- see also \cite{modin}, \cite{reich}  and reference therein.

Consider the ordinary differential equation \begin{equation}\label{ode}\frac{d}{dt}y(t)=X(y(t)),\end{equation} with $X$ a complete vector field on a manifold $M$ and $y(t)\in M$. The flow map for $X$ is denoted by $R_X:\mathbb{R}\times M\to M$. We use the notation $R_{X}(t,q)$ or simply $R_{X,t}(q)$. The flow $R_{X,t}$ may be expressed  using  a exponential map notation as $R_{X,t}(q)=\exp(tX)(q)$, where $t$ is a parameter and $\exp:\mathfrak{X}(M)\to\hbox{Diff}(M)$, with $\hbox{Diff}(M)$ denoting the set of diffeomorphisms on $M$ and $\mathfrak{X}(M)$ the set of vector fields on $M$. In the following, we assume that the flow $\exp(tX)$ is not explicitly integrable, and therefore one may use a numerical method to simulate the flow. Under this assumption, a numerical approximation to the solution of \eqref{ode} can be given by constructing a family of diffeomorphisms $\{\Phi_h\}_{h\geq 0}$ and then, for each $h$ fixed, it may be possible to obtain the sequence $\{q_{h,n}\}_{n\in\mathbb{N}}$ satisfying $\Phi_h(q_{h,n})=q_{h,n+1}$, called a numerical integrator. 
A \textbf{numerical integrator} for $X$ is a family of one-parameter diffeomorphisms $\Phi_{h}:M\to M$ (smooth in $h$) satisfying $\Phi_{0}(x)=x$ with $x\in M$, and $\Phi_{h}(x)-\exp(hX)(x)=\mathcal{O}(h^{p+1})$ with $p\geq 1$ being the order of the integrator.
Let us consider now the special case when $M=T^{*}Q$ (as in this paper).
We recall that an integrator $\Phi_h$ is  \textbf{symplectic} if it is a symplectic diffeomorphism with respect to the symplectic canonical structure $\omega_Q$ on $T^{*}Q$ for each $h>0$.

 Consider the Hamilton equations \eqref{tdepham} for $H:\R\times T^{*}Q\to\R$, that is the integral curves of the evolution vector field  \begin{equation}\label{forcedvf}E_H=\frac{\partial}{\partial t}+\frac{\partial H}{\partial p}\frac{\partial}{\partial q}-\frac{\partial H}{\partial q}\frac{\partial}{\partial p}\end{equation} 

 We aim to study backward error analysis for $\widetilde{\Psi}_d^{k,k+1}:T^{*}Q\to T^{*}Q$, the discrete Hamiltonian flow defined in Definition \ref{defhamin} for the non-autonomous Hamiltonian system \eqref{tdepham} at each fixed $t\in\R$ - recall that $\Psi_{t,s}:\mathcal{U}_t\subseteq T^*Q\to T^*Q$ is a symplectomorphism, in particular for $s=h$.

Using the extended Hamiltonian $H^{ext}: T^*({\mathbb R}\times Q)\rightarrow {\mathbb R}$ defined by
\[
H^{ext}(t,q, \mu, p)=\mu+H(t, q,p),
\]
the corresponding equations of motion for the Hamiltonian vector field $X_{H^{ext}}$ are
\begin{eqnarray*}
\dot{q}&=&\frac{\partial H^{ext}}{\partial p}=\frac{\partial H}{\partial p},\\
\dot{p}&=&-\frac{\partial H^{ext}}{\partial q}=-\frac{\partial H}{\partial q},\\
\dot{t}&=&\frac{\partial H^{ext}}{\partial \mu}=1,\\
\dot{\mu}&=&-\frac{\partial H^{ext}}{\partial t}=-\frac{\partial H}{\partial t}.
\end{eqnarray*}
The Hamiltonian $X_{H^{ext}}$ projects onto 
$E_H$ and therefore also their  flows are related by the projection ${pr}: T^*({\mathbb R}\times Q)\rightarrow {\mathbb R}\times T^*Q$ given by ${pr} (t, q, \mu, p)=(t, q, p)$. 

Now we will see how to naturally  extend the flow $\widetilde{\Psi}_d^{k,k+1}: T^*Q\rightarrow T^*Q$ to a symplectic discrete flow $\widetilde{\Psi}_h^{ext}: T^*({\mathbb R}\times Q)\rightarrow T^*({\mathbb R}\times Q)$. Consider the extended discrete Lagrangian $L^{ext}_d:  Q\times Q\times {\mathbb R}\times  {\mathbb R} \rightarrow {\mathbb R}$ subjected to the  constraint $t_{k+1}=t_k+h$ and $t_0=0$  then 
\[
L^{ext}_d ( q_k, q_{k+1},t_k, t_{k+1})=L^k_d(q_k, q_{k+1}),
\]
for  $t_k=kh$ and $t_{k+1}=h(k+1)$. 
Applying discrete variational calculus subjected to  constraints we obtain the following implicit symplectic method (see \cite{mawest, stern, Diaz}) 
\begin{eqnarray*}
p_k&=&- D_1 L^{ext}_d ( q_k, q_{k+1},t_k, t_{k+1})=-D_1L^k_d(q_k, q_{k+1}),\\
p_{k+1}&=&D_2 L^{ext}_d ( q_k, q_{k+1},t_k, t_{k+1})=D_2L^k_d (q_k, q_{k+1}),\\
\mu_k&=&- D_3L^{ext}_d ( q_k, q_{k+1},t_k, t_{k+1})+\lambda_k,\\
\mu_{k+1}&=&D_4L^{ext}_d ( q_k, q_{k+1},t_k, t_{k+1})+\lambda_k,\\
t_{k+1}&=& t_k+h,
\end{eqnarray*}
where $\lambda_k$ is a Lagrange multiplier associated to the  constraint $t_{k+1}=t_k+h$. 
These equations implicitly define a symplectic flow  $
{\widetilde{\Psi}_h}^{ext}: T^*( {\mathbb R}\times Q)\rightarrow T^*({\mathbb R}\times Q)$ by
\[
{\widetilde{\Psi}_h}^{ext} (t_k, q_k, \mu_k, p_k)= (t_k+h, q_{k+1},  \mu_{k+1}, p_{k+1}).
\]
Moreover  $
{\widetilde{\Psi}_h}^{ext}$  it is a numerical integrator for $X_{H^{ext}}$

Applying classical results of backward error analysis \cite{hairer, hansen} we can derive a modified 
 Hamiltonian  vector field $\bar{X}_{H^{ext}}$ that can be written as an asymptotic expansion in terms of the step-size $h>0$ as \begin{equation}\label{asymptotic} \bar{X}_{H^{ext}}=\sum_{r=0}^{\infty}h^{r}X_r,\end{equation} where each $X_r$ is a real analytic vector field on $T^{*}({\mathbb R}\times Q$) and it may be determined  by the integrator $
  {\widetilde{\Psi}_h}^{ext}$  as \begin{equation}\label{limit}X_r (t,q, \mu, p)=\lim_{h\to 0}\frac{
  	{\widetilde{\Psi}_h}^{ext}(t,q, \mu, p)-\exp(hX_{h,r-1})(t,q, \mu, p)}{h^{r}},\end{equation} with  $X_0=X_{H^{ext}}$ and $\displaystyle{X_{h,r}:=\sum_{j=0}^{r}h^{j}X_j}$.  

Since the  discretization ${\widetilde{\Psi}_h}^{ext}$ is symplectic there exist   functions
$H^{ext}_r: T^*( {\mathbb R}\times Q)\rightarrow {\mathbb R}$
such that each $X_r=X_{H^{ext}_r}$ with $X_0=X_{H^{ext}}$ \cite{hairer}. That is, the modified vector field 
$\bar{X}_{H^{ext}}$ associated to $
{\widetilde{\Psi}_h}^{ext}$ is Hamiltonian $\bar{H}_{ext}: T^*({\mathbb R}\times Q)\rightarrow {\mathbb R}$ with Hamiltonian function with formal expansion 
\[
\bar{H}_{ext}=H_{ext}+\sum_{r=1}^{\infty}h^{r} H^{ext}_r.
\]
 Furthermore,
because the equation of motion in the variable $t$  is integrated exactly (that is, $t_{k+1}=t_k+h$)
we have  that 
$\bar{H}_{ext}(q,t,p, \mu)=\mu+\bar{H}(q, p, t)$
and, in consequence, also $H^{ext}_r: {\mathbb R}\times T^*Q\rightarrow {\mathbb R}$.  
We can consider the truncated  Hamiltonians:
 $\bar{H}^N_{ext}=H_{ext}+\displaystyle{\sum_{r=1}^{N}h^{r }H^{ext}_r}$.
	Therefore we have a	truncated Hamiltonian 
	$\bar{H}^N=H+\displaystyle{\sum_{r=1}^{N}h^{r }H^{ext}_r}$ on ${\mathbb R}\times T^*Q$.
 We have corresponding  \textbf{evolution vector field} \(E_{\bar H^N}\in \mathfrak{X}(T^*Q\times \R)\) determined by
\begin{equation}
\label{eq:qqq-1}
i_{E_{\bar H^N}}\Omega_{\bar H^N}=0\; ,\qquad i_{{E_{\bar H^N}}}\eta=1
\end{equation}
As a consequence its flow preserves the 2-form $\Omega_{\bar H^N}$ and the 1-form $\eta$, being two important properties of this type of geometric integrators. 
In local coordinates the evolution vector field is given by 
\[
{E_{\bar H^N}}=\frac{\partial}{\partial t}+\frac{\partial {\bar H^N}}{\partial p_i}\frac{\partial}{\partial q^i}-\frac{\partial {\bar H^N}}{\partial q^i}\frac{\partial}{\partial p_i}.
\]
As in Section \ref{sec: ham} from the flow of ${E_{\bar H^N}}$ we induce the two-parameter symplectic family of maps $\Psi_{t,s}^{E_{\bar H^N}}:  T^* Q\rightarrow T^*Q$. 
	
From our previous considerations we deduce that 
 $\widetilde{\Psi}_d^{k,k+1}(q,p)-\Psi_{kh,h}^{E_{\bar H^N}}(q,p)=\mathcal{O}(h^{N+1})$.  

In particular one has the following result for autonomous systems from \cite{hansen} adapted to our non-autonomous context.

\begin{lemma}\label{hansen}[Adapted from A. C. Hansen (2011) Theorem $4.1$ \cite{hansen}]
	Let $T^{*}Q$ be a real and analytic smooth manifold, $\hbox{d}$ a Riemannian distance  on $T^{*}Q$,  a real analytic evolution vector field $E_{\bar H^N}$ on $T^{*}Q$  and $\widetilde{\Psi}_d^{k,k+1}$ be a numerical integrator deduced from a family of discrete Lagrangians $\{L_d^k\}$ 
 such that the induced symplectic method $\widetilde{\Psi}_d^{k,k+1}: T^*Q\rightarrow T^*Q$ is of order $p$ such that it is  analytical and $(q, p)\in\mathcal{K}\subset T^{*}Q$ with $\mathcal{K}$ compact. For each time step $k$ there exists $\tau\in\mathbb{Z}$ depending on $h$ and positive constants $C,\alpha,\gamma$ such that for $\Psi_{t,s}^{E_{\bar H^{\tau}}}:  T^* Q\rightarrow T^*Q$ such that $\displaystyle{\hbox{d}\left(\widetilde{\Psi}_d^{k,k+1}(q, p),\Psi_{kh,h}^{E_{\bar H^N}}(q,p)\right)\leq Che^{-\gamma/h}}$ for all $(q,p)\in\mathcal{K}$ and $h\leq \alpha$, where $\widetilde{\Psi}_d^{k,k+1}$ must be considered as $\widetilde{\Psi}_d^{k,k+1}:=\varphi\circ\widetilde{\Psi}_d^{k,k+1}\circ\varphi^{-1}$ for a given local chart $(U,\varphi)$ on $T^{*}Q$.
\end{lemma}

Finally, consider the truncated  Hamiltonian $\bar{H}^N_{ext}=H_{ext}+\displaystyle{\sum_{r=p}^{N}h^{r }H^{ext}_r}$.
Following \cite{hairer}, Section IX.8  we obtain the following result: 

\begin{theorem}\label{thorder}
	Assume that   the Hamiltonian function $H_{ext} : \mathcal{U}\subset T^*({\mathbb R}\times Q)\rightarrow {\mathbb R}$  where $\mathcal{U}$ is an open subset,  
	and apply the symplectic method $
	{\widetilde{\Psi}_h}^{ext}$. If
	the numerical solution stays in a compact set $\mathcal{K}\subset \mathcal{U}$, then there exist $h_0$  and
	$N = N(h)$, ($N$ equal to the
largest integer satisfying $hN \leq  h_0$)  such that
	\begin{eqnarray*}
		\bar{H}^N_{ext} (q_k, t_k, p_k, \mu_k)
	&=& \bar{H}^N_{ext} (q_0, t_0, p_0, \mu_0)
+	{\mathcal O}(e^{
	-h_0/2h}),\\
		{H}_{ext} (q_k, t_k, p_k, \mu_k)
	&=& {H}_{ext} (q_0, t_0, p_0, \mu_0) + {\mathcal O}(h^p),
	\end{eqnarray*}
	over exponentially long time intervals $nh \leq  e^{h_0/2h}$.
	\end{theorem}

\section{Application to formation control of double integrator agents}\label{Sec: formationcontrol}

Formation control of agents with double integrator dynamics can be seen as a stabilization system whose evolution can be described by a time-dependent Lagrangian function. Next we employ the previous constructions on variational integrators for time-dependent Lagrangian systems with symmetries and backward error analysis in the context of distance-based formation control algorithms.

\subsection{Double integrator formation stabilization systems}

Consider $n\geq 2$ autonomous agents whose positions are denoted by $q_i\in\R^{d}$, $d=\{2,3\}$ and denote by $q\in\R^{dn}$ the stacked vector of agents' positions. Agent's evolve under a double integrator dynamics, that is 
$\displaystyle{	\begin{cases}
	\dot{q} =v \\
	\dot{v} = u
	\end{cases}}$.

The neighbor relationships between agents are described by an undirected graph $\mathbb{G} = (\mathcal{N}, \mathcal{E})$ with the ordered edge set $\mathcal{E}\subseteq\mathcal{N}\times\mathcal{N}$. The set of neighbors for $i\in\mathcal{N}$, denoted by $\mathcal{N}_i$, is defined by $\mathcal{N}_i:=\{j\in\mathcal{N}:(i,j)\in\mathcal{E}\}$. Agents can sense the relative positions of its nearest neighbors, in particular, agents can measure its Euclidean distance from other agents in the subset $\mathcal{N}_i \subseteq \mathcal{N}$. We define the elements of the incidence matrix $B\in\R^{|\mathcal{N}|\times|\mathcal{E}|}$ that establish the neighbors' relationships for $\mathbb{G}$ by $b_{i,w} = \begin{small}\begin{cases}+1  &\text{if} \quad i = {\mathcal{E}_w^{\text{tail}}} \\
		-1  &\text{if} \quad i = {\mathcal{E}_w^{\text{head}}} \\
		0  &\text{otherwise}
	\end{cases}\end{small}$, where $\mathcal{E}_w^{\text{tail}}$ and $\mathcal{E}_w^{\text{head}}$ denote the tail and head nodes, respectively, of the edge $\mathcal{E}_w$, i.e., $\mathcal{E}_w = (\mathcal{E}_w^{\text{tail}},\mathcal{E}_w^{\text{head}})$. The stacked vector of relative positions between neighboring agents, denoted by $z\in\R^{d|\mathcal{N}|}$, is given by $z = \overline B^T q$, where $\overline B := B \otimes I_{d}\in\R^{d|\mathcal{N}|\times d|\mathcal{E}|}$ with $I_{d}$ being the $(d\times d)$ identity matrix, and $\otimes$ the Kronecker product. Note that $z_w \in \R^d$ and $z_{w+|\mathcal{E}|}\in\R^d$ in $z$ correspond to $q_i - q_j$ and $q_j - q_i$ for the edge $\mathcal{E}_w$. 

We consider the desired distance between neighboring agents over the edge $\mathcal{E}_w$ as $d_{w}\in\mathbb{R}$ and we further define the squared distance error for the edge $\mathcal{E}_w$ as $e_{w}=\left\|q_{i}-q_{j}\right\|^{2}-d_{w}^{2}=\left\|z_{w}\right\|^{2}-d_{w}^{2}$, with the stacked squared distance vector error denoted by $e = \left[e_{1}, \ldots, e_{|\mathcal{E}|}\right]^{\top}$. For $w\in\{1,\ldots,|\mathcal{E}|\}$, the set of desired shapes is defined by $\mathcal{S}=\{z\in\R^{d|\mathcal{N|}}|\,\,||z_w||=d_w \}$.

 A \textit{framework} for $\mathbb{G}$ is then defined as the pair $(\mathbb{G},q)$. In this work, conditions to guarantee convergence to desired formations are based on the property called \textit{rigidity} of the desired formation shape. According to this, the \textit{rigidity matrix} for the framework~$(\mathbb{G},p)$ is defined as (see \cite{asimov} for instance)
$\displaystyle{R(z)=\frac{1}{2} \frac{\partial \ell_{\mathbb{G}}(p)}  {\partial p} = D(z)^{\top} \overline{B} \in \R^{|\mathcal{E}| \times d|\mathcal{N}|}}$, with~$D(z) = \hbox{diag}(z_{1}, \ldots,z_{|\mathcal{E}|}) \in \R^{d|\mathcal{E}| \times |\mathcal{E}|}$ and distance measure function~$\ell_{\mathbb{G}}: \R^{d|\mathcal{N}|} \rightarrow \R^{|\mathcal{E}|}$ defined by 
$\ell_{\mathbb{G}}(p)=\left(\left\|p_{i}-p_{j}\right\|^{2}\right)_{(i, j) \in \mathcal{E}}=D^\top(z)z$.

 A framework~$(\mathbb{G},q)$ is said to be \textit{rigid} if it is not possible to smoothly move one node of the framework without moving the rest while maintaining the inter-agent distance given by~$\ell_{\mathbb{G}}(q)$, see~\cite{asimov}. An \textit{infinitesimally rigid} framework is a rigid framework which is invariant under and only under infinitesimally transformations under~$R(z)$, i.e.,~$\ell_{\mathbb{G}}(q+\delta q)=\ell_{\mathbb{G}}(q)$ where~$\delta q$ denotes an infinitesimal displacement of~$q$. It is well known (see \cite{asimov}) that a framework ~$(\mathbb{G},q)$ is infinitesimally rigid in~$\R^d$ if~$q$ is a regular value of ~$\ell_{\mathbb{G}}(q)$ and~$(\mathbb{G},q)$ is rigid in~$\R^d$. In particular,~$(\mathbb{G},q)$ is \textit{infinitesimally rigid} in~$\R^2$ if~$\hbox{rank} R(z)=2n-3$ (respectively, ~$\hbox{rank} R(z)=3n-6$ in~$\R^3$). The framework~$(\mathbb{G},q)$ is said to be \textit{minimally rigid} if it has exactly~$2n-3$ edges in~$\R^2$ or~$3n-6$ edges in~$\R^{3}$. This means that if we remove one edge from a minimally rigid framework~$(\mathbb{G},q)$, then it is not rigid anymore. Thus, the only motions over the agents in a minimally rigid framework, while they are already in the desired shape, are the ones defining translations and rotations of the shape, see~\cite{oh2015survey}. Along the remained of the paper we assume that the framework $(\mathbb{G},q)$ is infinitesimally and minimally rigid.

By considering the control law $u(t)=-\mathcal{K}v-R^\top(z)e(z)$, the closed loop system is given by 
\begin{align}\label{formation}
	\begin{cases}
	\dot{q} =v \\
	\dot{v} = -\mathcal{K}v-R^\top(z)e(z).
	\end{cases}
\end{align} with $R$ being the rigidity matrix for $\mathbb{G}$, $\mathcal{K}=K\otimes I_{d}$ and $K$ the gain diagonal matrix with the $i$-th entry being $k_i>0$. The closed-loop system \eqref{formation} is called double integrator formation stabilization system (see \cite{oh2015survey} for instance). Note that the role of equations \eqref{formation} is to stabilize a desired infinitesimal and minimal rigid shape and reach a stationary formation with zero velocities of the agents. 

To reach the desired shape $\mathcal{S}$, for each edge $\mathcal{E}_w=(i,j)$, in the infinitesimally and minimally rigid framework, one introduces the artificial potential functions $V_{w}:\R^{d}\to\R$, given by
\begin{align}
	V_{w}(z_w)=\frac{1}{4}(||z_{w}||^2-d_{w}^2)^2,
	\label{eq: Vij}
\end{align} to provide a measure for the interaction between agents and their nearest neighbors (see \cite{oh2015survey} for a detailed discussion on the choices of elastic potential functions).  In these potentials,~$z_w$ denotes the relative position between agents for the edge $\mathcal{E}_w$, and $d_{w}$ denotes the desired length for the edge $\mathcal{E}_w$. Note also that the artificial potential (\ref{eq: Vij}) is not unique, and it can be given by other similar expressions as it was discussed by~\cite{oh2015survey}. Therefore, we can define the artificial potential function $V_{ij}:\R^{d|\mathcal{N}|}\to\R$ for the overall networked control system as $V_{ij}(z)=\sum_{w=1}^{|\mathcal{E}|}V_w(z_w)$. With this notation, the system \eqref{formation} can be written as \begin{align}\label{formation2}
	\begin{cases}
	\dot{q} =v \\
	\dot{v} = -\mathcal{K}v-\nabla_{q}V_{ij}.
	\end{cases}
\end{align} that is, for each $i\in\mathcal{N}$, the system \eqref{formation2} can be expressed as 
\begin{align}\label{formation3}
	\begin{cases}
	\dot{q}_i =v_i \\
	\dot{v}_i = -\kappa_iv_i-\nabla_{q_i}V_{ij}.
	\end{cases}
\end{align} with $\kappa_i>0$ being the diagonal elements of the matrix $\mathcal{K}$ which, without loss of generalities, from now on, we will assume that $\kappa:=\kappa_i=\kappa_j>0$ for all $i,j$.

Note that the double integrator formation stabilization system \eqref{formation3} can be given by the Euler-Lagrange equations for the time-dependent Lagrangian function $L:\mathbb{R}\times\mathbb{R}^{nd}\times\mathbb{R}^{nd}\to\mathbb{R}$ given by  \begin{equation}\label{stabletimedependet}L(t,q,\dot{q})=e^{\kappa t}\left(\frac{1}{2}\sum_{i=1}^{n}||\dot{q}
_i||^{2}-V_{ij}(q)\right).\end{equation}

\subsection{Derivation of the discretized equations of motion}

To construct the geometric integrator, the velocities for each agent $i\in\mathcal{N}$ are discretized by finite-differences, i.e., $\displaystyle{\dot{q}_i=\frac{q_{k+1}^i-q_k^i}{h}}$ for $t\in[t_k,t_{k+1}]$. The discrete Lagrangian $L_{d,h}^{k}:\mathbb{R}^{d|\mathcal{N}|}\times\mathbb{R}^{d|\mathcal{N}|}\to\mathbb{R}$ is given by setting the trapezoidal discretization for the time-dependent Lagrangian $L$ given by \eqref{stabletimedependet}, that is, \begin{align*}
L_{d,h}^{k}(q_k,q_{k+1})=\frac{h}{2}L\left(kh, q_k,\frac{q_{k+1}-q_k}{h}\right)+\frac{h}{2}L\left((k+1)h,q_{k+1},\frac{q_{k+1}-q_k}{h}\right)\end{align*}where, $h>0$ is the time step. 

The discrete Euler-Lagrange equations for $L_{d,h}^{k}$ are given by \begin{align}
    0=&(q_{k+1}-q_k)e^{\kappa(kh)}-(q_{k+2}-q_{k+1})e^{\kappa h(k+2)}\label{deleq}\\
&-e^{\kappa h(k+1)}(q_k-2q_{k+1}+q_{k+2}+2h^2\nabla_{q_{k+1}}V_{ij}^{d}(q_{k+1}^{i},q_{k+1}^{j}))\nonumber,
\end{align} where we have used that $$V_{ij}^{d}=\frac{h}{2}\sum_{j\in\mathcal{N}_i}(V_{ij}^d(q_k^{i},q_k^j)+V_{ij}^d(q_{k+1}^i,q_{k+1}^j)),$$ that is, $V_{ij}^{d}$ is the trapezoidal discretization of $V_{ij}$.

After some calculus we can write equations \eqref{deleq} as the following explicit integration scheme \begin{equation}\label{integrator-explicit}
    q_{k+2}=\hat{\kappa}_{h}q_{k+1}-\kappa_{h}q_k-\bar{\kappa}_{h}\nabla_{q_{k+1}}V_{ij}^{d}(q_{k+1}^{i},q_{k+1}^{j}),
\end{equation}with $\kappa_{h}=\frac{1+e^{-\kappa h}}{1+e^{\kappa h}}$, $\bar{\kappa}_{h}=\frac{2h^{2}}{1+e^{\kappa h}}$, $\hat{\kappa}_{h}=\frac{2+e^{\kappa h}+e^{-\kappa h}}{1+e^{\kappa h}}$, that is, for each agent $i\in\mathcal{N}$ \begin{align*}	\begin{cases}
    q_{k+2}^{i}&=\hat{\kappa}_hq_{k+1}^{i}-\kappa q_{k}^{i}-\bar{\kappa}_h\displaystyle{\sum_{j\in\mathcal{N}_i}}
    \Gamma_{ij}^{k}(q_{k+1}^{i}-q_{k+1}^{j}),\\
    \Gamma_{ij}^{k}&=(q_{k+1}^{i}-q_{k+1}^j)-d_{ij}^{2}.\end{cases}
\end{align*}

Note that the previous equations are a set of $d|\mathcal{N}|(N-1)$ for the $d|\mathcal{N}|(N+1)$ unknowns $\{q_k^i\}_{k=0}^{N}$, with $1\leq i\leq n=|\mathcal{N}|$. Nevertheless the boundary conditions on initial positions and velocities of the agents $q_0^{i}=q_i(0)$, $v_{q_0}^i=\dot{q}_i(0)$ contribute to $2dn$ extra equations that convert eqs. \eqref{deleq} in a nonlinear root finding problem of $dn(N-1)$ equations and the same amount of unknowns. To start the algorithm we use the boundary conditions for the first two steps, that is, $q_0^i=q_i(0)$ and $q_1^i=hv_{q_0}^i+q_{0}^i=h\dot{q}_{i}(0)+q_{i}(0).$

\begin{remark}Observe also that the Lagrangian $L_{d,h}^{k}$ is $SE(d)$-invariant, since the inter-agent potential is $SE(d)$-invariant, therefore applying the discrete Noether Theorem \ref{noetherd}. Both the linear and angular momentum in double-integrator formation
systems are related with steering
controller design for coordinating a formation as a whole at the steady state by using the linear and angular momentum of the centroid and therefore the variational integrators developed in this work could be used as for the steering control to achieve a desired formation.%
 \end{remark}

\subsection{Simulation results}
\begin{figure}[h!]
\centering
\includegraphics[scale=0.8]{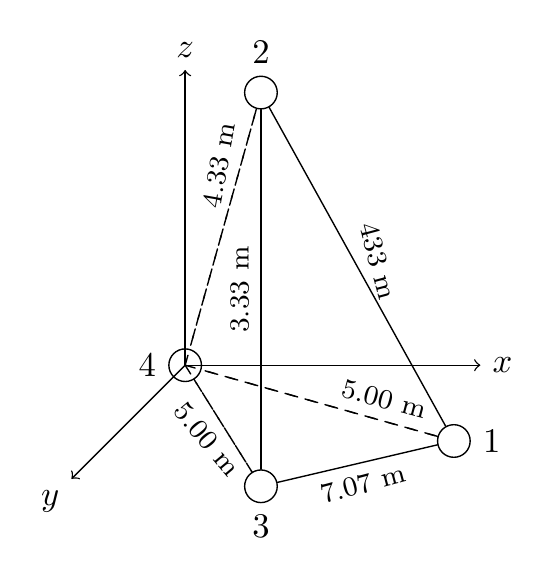}

\label{f2}

\caption{Infinitesimal and minimal rigid graph considered in the numerical simulations.}\label{f1}
\end{figure}

Next, for simulation purposes we willrestrict ourselves to the case $n=4$, $d=3$, where the desired formation shape is depicted in Figure \ref{f1} with neighbour relationships given by $\mathcal{N}_1=\{2,3,4\}$, $\mathcal{N}_2=\{1,3,4\}$, $\mathcal{N}_3=\{1,2,4\}$ and $\mathcal{N}_4=\{1,2,3\}$.

 The explicit integration scheme \eqref{integrator-explicit} is given by  \begin{equation}
	\begin{cases}
q_{k+2}^1=&G_{k}^1-\bar{\kappa}_h\left(\Gamma_{12}^{k+1}(q_{k+1}^1-q_{k+1}^2)\right.
\left.+\Gamma_{13}^{k+1}(q_{k+1}^1-q_{k+1}^3)+\Gamma_{14}^{k+1}(q_{k+1}^1-q_{k+1}^4)\right)\\
q_{k+2}^2=&G_{k}^2-\bar{\kappa}_h\left(\Gamma_{21}^{k+1}(q_{k+1}^2-q_{k+1}^1)+\Gamma_{23}^{k+1}(q_{k+1}^2-q_{k+1}^3)+\Gamma_{24}^{k+1}(q_{k+1}^2-q_{k+1}^4)\right)\\
q_{k+2}^3=&G_{k}^3-\bar{\kappa}_h\left(\Gamma_{31}^{k+1}(q_{k+1}^3-q_{k+1}^1)\right.\left.+\Gamma_{32}^{k+1}(q_{k+1}^3-q_{k+1}^2)+\Gamma_{34}^{k+1}(q_{k+1}^3-q_{k+1}^4)\right)\\
q_{k+2}^4=&G_{k}^4-\bar{\kappa}_h\left(\Gamma_{41}^{k+1}(q_{k+1}^4-q_{k+1}^1)+\Gamma_{42}^{k+1}(q_{k+1}^4-q_{k+1}^2)+\Gamma_{43}^{k+1}(q_{k+1}^4-q_{k+1}^3)\right)
	\end{cases}
	\label{eq: dyndis2}
\end{equation}
where  $G_{k}^i=G(q_{k}^i,q_{k+1}^i)=\hat{\kappa}_hq_{k+1}^{i}-\kappa_hq_k^i$, $q_k^i=(x_k^i,y_k^i, z_k^i)\in\R^3$, $i=1,\ldots,4$.

\begin{figure}[H]
\centering
\parbox{5cm}{
\includegraphics[width=0.47\columnwidth]{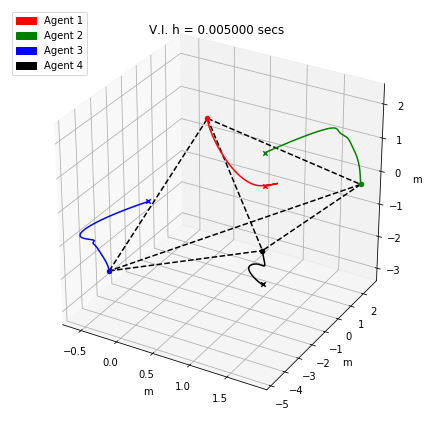}

}
\qquad\qquad\qquad\qquad\qquad
\begin{minipage}{5cm}
\includegraphics[width=1.25\columnwidth]{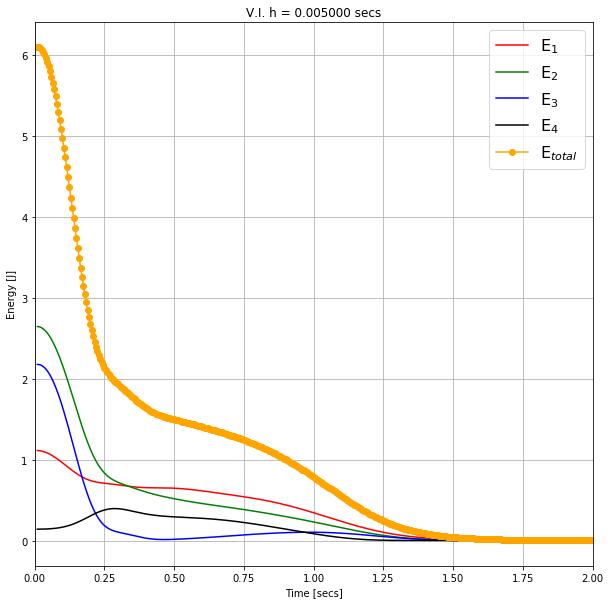}

\end{minipage}\caption{Convergence of agents' trajectories by employing the variational integrator (left) and evolution of the discrete energy function in the 3-dimensional formation with piramidal shape (right). The crosses denote the initial positions.}\label{f5}
\end{figure}

Initial positions were $q_{0}=[1,0,0,1,0,1,0,-3,0,1,0,-3]$ and we set the initial velocities to zero and damping gains $\kappa=13$. In this case, an end time was settled of $2$ seconds in steps of $h=0.005$ seconds, resulting in $N=400$ iterations. In Figure \ref{f5} on the left we show the convergence of agents' trajectories by employing the variational integrator and in Figure \ref{f5} on the right we shows the decrease of the energy, both per agent and total.

\label{sec: sim}

The energy function was discretized using a trapezoidal discretization.  In particular, the total energy of each agent $E_i:TQ\to \mathbb{R}$ is given by $$E_i(q_i,\dot{q}_i)=\frac{1}{2}||\dot{q}_i||^2+\frac{1}{2}\sum_{j\in\mathcal{N}_i}V_{ij}(q_i,q_j).$$ Using the trapezoidal rule for $E_i$, the discrete energy function for each agent $E_i^d:\mathbb{R}^n\times\mathbb{R}^n\to\mathbb{R}$ is given by \begin{equation}E_i^d(q_k^i,q_{k+1}^i)=\frac{1}{2h^2}(q_{k+1}^i-q_k^i)^2+\frac{1}{4}\sum_{j\in\mathcal{N}_i}(V_{ij}^d(q_k^{i},q_k^j)+V_{ij}^d(q_{k+1}^i,q_{k+1}^j)).\end{equation}

Note that the evolution of the system's energy presents a decay behavior with a relativelly fast decay rate. In particular this indicates that agents can employ the variational integrator for their estimation algorithms to save energy consumption since they have a lower computational cost than traditional numerical solutions and without compromising accuracy (the integrator is explicit as an Euler integrator). In fact, the accuracy in a simulation is also crucial when a multi-agent system can consists of a significant number of agents and links, i.e., the bigger the number of initial conditions, the bigger the sensitivity for the agents' trajectories. We compare the performance of the variational integrator (\ref{eq: dyndis2}) and the Euler discretization of (\ref{formation3}) since both methods are similar in terms of computational cost per time step and explicit. Indeed, other methods like Runge-Kutta can give excellent results in terms of accuracy. However, one needs to evaluate the differential equation (\ref{formation3}) several times per discrete step depending on the desired accuracy, hence increasing the computational cost. For the comparison, we consider four agents whose desired shape is defined from Figure \ref{f2}. 
\begin{figure}[h!]
\centering
\includegraphics[width=0.45\columnwidth]{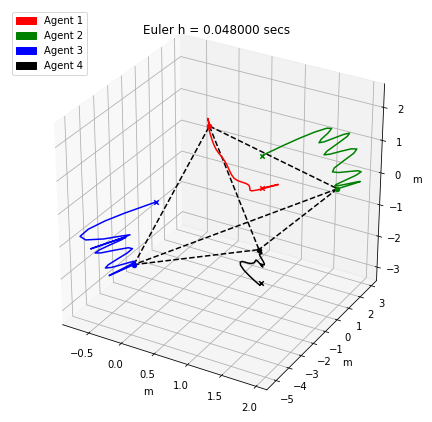}
\includegraphics[width=0.45\columnwidth]{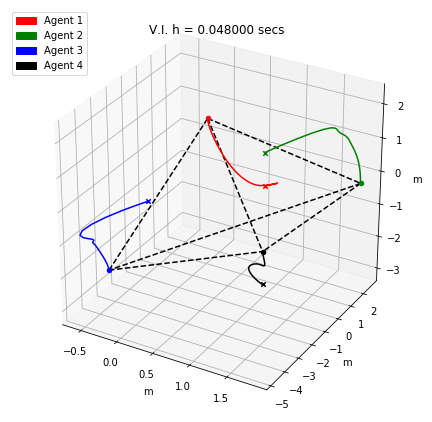}
\includegraphics[width=0.4\columnwidth]{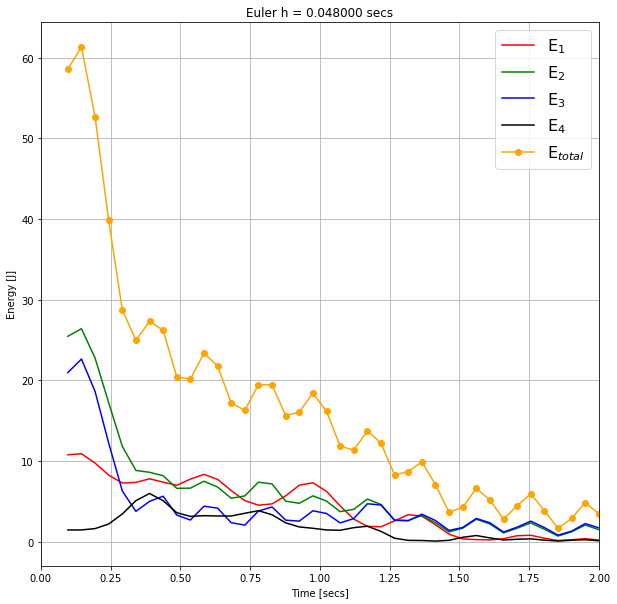}
\includegraphics[width=0.4\columnwidth]{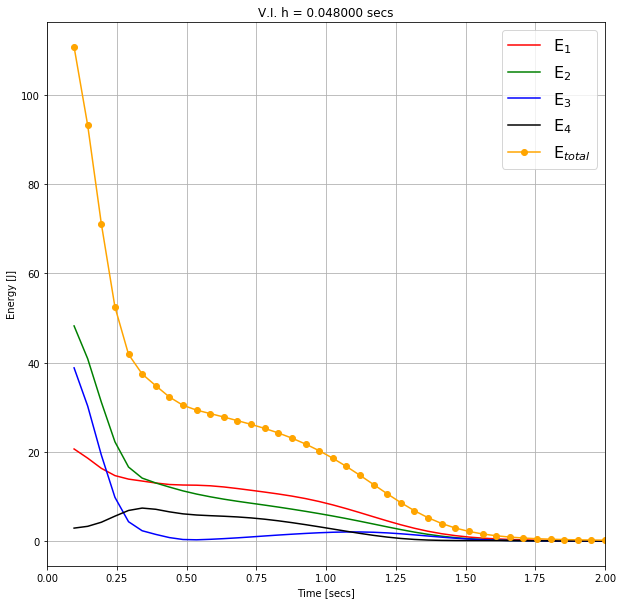}
\caption{While Euler method is stable for $h=0.05$, it is for $h=0.008$ or lower than the transitory of the agents' trajectories, and therefore the final desired squared shape, are consistent with the results from the variational integrator. The crosses denote the initial positions.}
\label{fig: sim2}
\end{figure}

While the Euler method starts to be stable, i.e., the solution does not diverge to infinity, at $h = 0.05$, it presents a smooth behavior once the time step is lower than $h = 0.008$. However, as it can be checked in Figures \ref{fig: sim2}, the transitory and final shapes are notably different. In addition, in Figure \ref{fig: sim3} we can appreciate how the variational integrator outperforms the Euler integrator in terms of energy dissipation. Note that  the variational integrator decreases to zero fasther than the Euler method. In particular, the variational integrator does that in time $t=1.5$sec whereas the euler method at $t=4$sec.

\begin{figure}[h!]
\centering
\includegraphics[width=0.4\columnwidth]{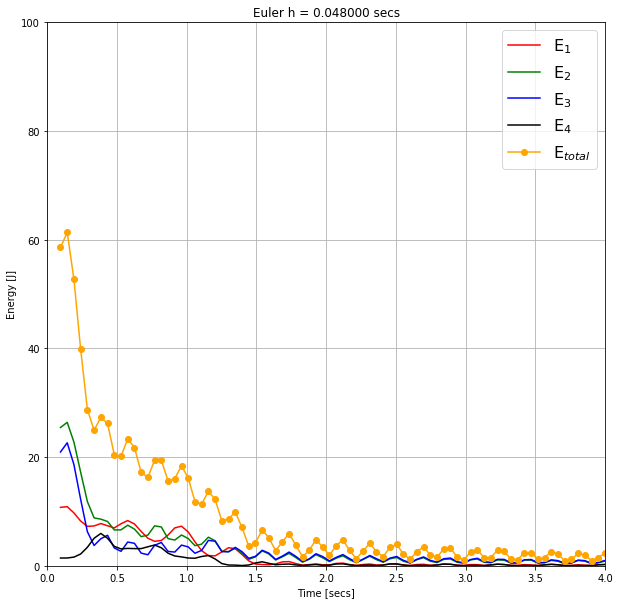}
\includegraphics[width=0.4\columnwidth]{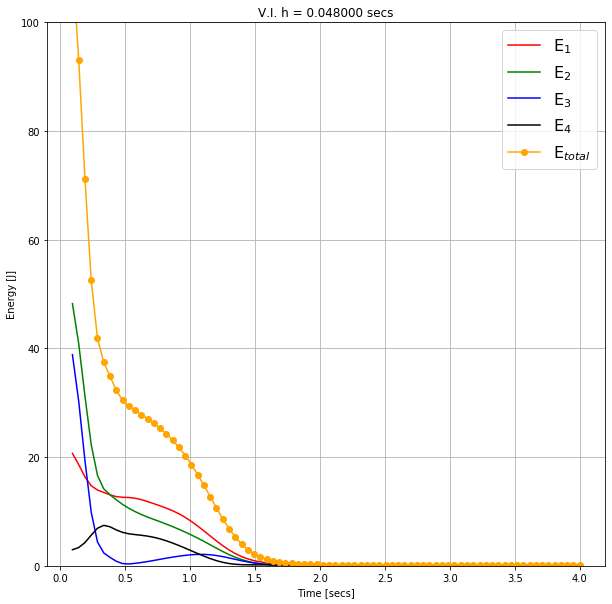}
\caption{Energy decay comparison between the variational integrator and the Euler integrator.}
\label{fig: sim3}
\end{figure}

\subsection{Application to the estimation of regions of attraction in formation control}\label{sec5.1}

Let us briefly review some concepts in formation control for the proposed numerical experiments. We define a desired configuration $q^*$ as a particular collection of fixed $q_i^*$ whose $SE(2)$-transformations define the \textit{desired shape}. Convergence results in (distance-based) formation control cover the local stabilization of the desired shape, and besides some analytical expressions for some particular cases of single-integrators, for double-integrator dynamics the neighborhoods or regions of attraction around $q^*$ (up to translations and rotations) are estimated numerically \cite{anderson, oh2015survey}. 

We say that two configurations $q^{1*}$ and $q^{2*}$ are \textit{congruent} if $||q^1_i - q^1_j|| = ||q^2_i - q^2_j||, i,j\in\mathcal{V}$ with $i\neq j$. Note that two configurations $q^{1*}$ and $q^{2*}$ can satisfy $||q^1_i - q^1_j|| = ||q^2_i - q^2_j||, (i,j)\in\mathcal{E}$ but might fail to be congruent, and therefore they \textit{do not describe the same shape}. We refer to the reader to the concept of \textit{rigidity} in formation-control \cite{anderson} on how to construct desired shapes from a set of desired distances between agents. Therefore we can have multiple shapes corresponding to a minimum of potential functions (\ref{eq: Vij}) in distance-based control. Obviously, the more edges in $\mathcal{E}$, the more constrains and fewer possible shapes given a collection of desired distances $d_{ij}$ with $(i,j)\in\mathcal{E}$. However, in practical scenarios we are interested in keeping a small number of edges, so the system is far from an \textit{all-to-all} scheme.

It is of crucial importance in robotic multi-agent systems to choose those initial conditions, or initial deployment, for the robots such that the eventual shape is congruent with the desired one. As we will illustrate, for agents that start at rest, i.e., with $\dot q_i(0) = 0$, some desired shapes have \textit{narrow} or even disconnected regions of attractions. We find such regions after intensive campaigns of numerical simulations where we are assisted by the variational integrators \eqref{integrator-explicit}. In particular, we will be able to run accurate simulations with significant large time steps with the same computational cost of a simple Euler integrator - \eqref{integrator-explicit} is an explicit integrator. The guarantees on the decreasing of the total energy of the system over time, together with a \textit{well behavior} of such energy evolution, is of vital importance due to the high sensitivity of the gradient of the potentials (\ref{eq: Vij}) to the positions of the agents, specially when they are far from the desired shape.

The following numerical experiment will estimate regions of attraction for desired shapes by exploiting the variational integrator \eqref{integrator-explicit}. In particular, we study the set of initial conditions for agent $i$ while the rest of agents are in the desired shape such that the eventual shape is congruent to the desired one. This case is common in practice for growing formations, and give us information on from which areas are \textit{safe} to deploy a new robot. In order to identify the region of attraction to the desired shape for one agent, we run $3000$ simulations with $hr = 5$, where $r$ is the number of steps and $h$ is the time step of the variational integrator \eqref{integrator-explicit}, and where we are also looking for those positions where the convergence time is lower than a threshold. In order to speed up the process for identifying the regions of attraction, we are interested in setting $h$ as big as possible for each simulation while having guarantees on the numerical stability, i.e., we are looking for $\alpha$ in Lemma \ref{hansen}. We can give the following expression for $\alpha$ (see Theorem $8.1$ and Example $8.2$, Section IX.8, pp. $367$ in \cite{hairer})%
\begin{equation}
	\alpha= \frac{R}{cM},\quad
	||f(q,p)|| \leq M,\quad
	||(q,p) - (q_0,p_0)||< 2R, \nonumber
\end{equation}
where $(q_0,p_0) \in \mathcal{K}:= \{(q,p) \in \mathbb{R}^{2n}\ \, \hbox{s.t.} \, ||p|| < c\}$, so for a fixed $c,R\in\mathbb{R}^{+}$ we can give $M$ from (\ref{formation3}) as follows
\begin{align}
	&||f(q,p)||^2 = ||c||^2 + \sum_{i=1}^{|\mathcal{V}|}\sum_{j\in\mathcal{N}_i} ||-\kappa p_i - \nabla V_{ij}(q_{ij})||^2 \leq ||c||^2 + 2|\mathcal{V}|\kappa^2 ||c||^2 + 4 \sum_{(i,j)\in\mathcal{E}} ||\nabla V_{ij}(q_{ij})||^2 \nonumber \\
	&\leq (1 + 2|\mathcal{V}|\kappa^2)||c||^2 + 4 |\mathcal{E}| \left(\operatorname{max}_{(i,j)\in\mathcal{E}}\left\{||q|| (|\,||q_{ij}||^2 - d_{ij}^2 \,|) \right\} \right)^2 \nonumber \\
	&\leq \begin{cases}
		(1 + 2|\mathcal{V}|\kappa^2)||c||^2 + 64 |\mathcal{E}| R^6, \, \text{if} \, ||q_{ij}||^2 > d_{ij}^2, \\
				(1 + 2|\mathcal{V}|\kappa^2)||c||^2 + 64 |\mathcal{E}| R^2 \operatorname{max}\{d_{ij}^4\},\,\text{if} \, d_{ij}^2 > ||q_{ij}||^2, 
	\end{cases} \nonumber
\end{align} for $q_{ij} \in \mathcal{K},\, (i,j)\in\mathcal{E}$.

For example, in our experiment with $\kappa = 13$, $|\mathcal{E}| = 9$ and $|\mathcal{V}| = 6$, then for initial conditions set by $c = R = 1$ where all the agents start with $\dot p_i(0) = 0$ we have that $\alpha = 0.014$. Then, we have chosen $h = 0.014$, and with the required initial conditions, we have observed that with $200$ steps, the agents have enough time to converge to an equilibrium. To determine whether an eventual shape in a simulation is congruent to the desired one we check if the discrepancy of distances between agents in their final positions is lower than $1\%$ with respect to the desired shape in $q^*$. Indeed, we also check that the eventual velocities for the agents are also close enough to zero, e.g., $||\dot p_i(T)|| < 0.1$, being $T$ the final time of the simulation. Figure \ref{fig: sim3} show the results on regions of attraction for a desired  infinitesimal and minimal rigid shape when all the agents excepting one start at the desired shape. 

\begin{figure}[h!]
\centering

\includegraphics[width=0.49\columnwidth]{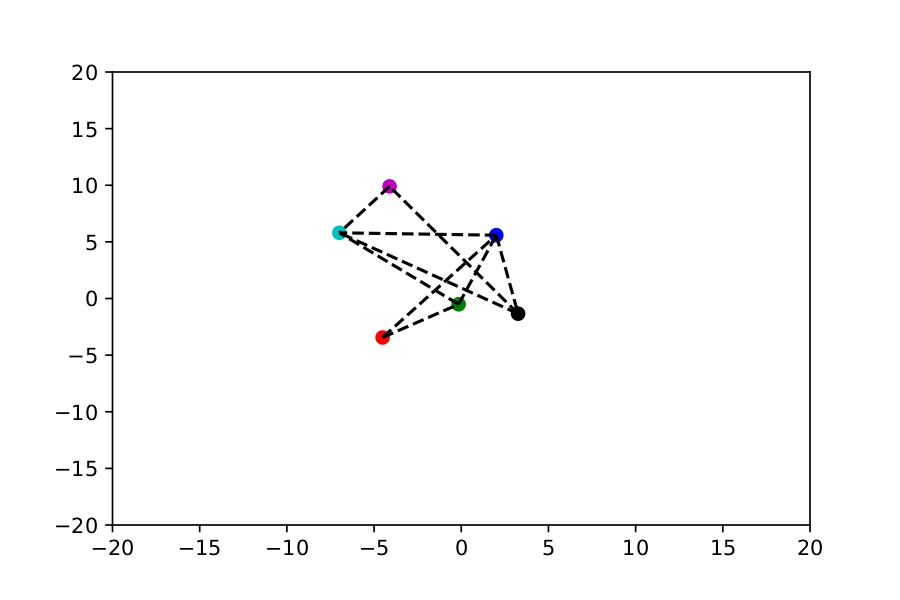}
\includegraphics[width=0.49\columnwidth]{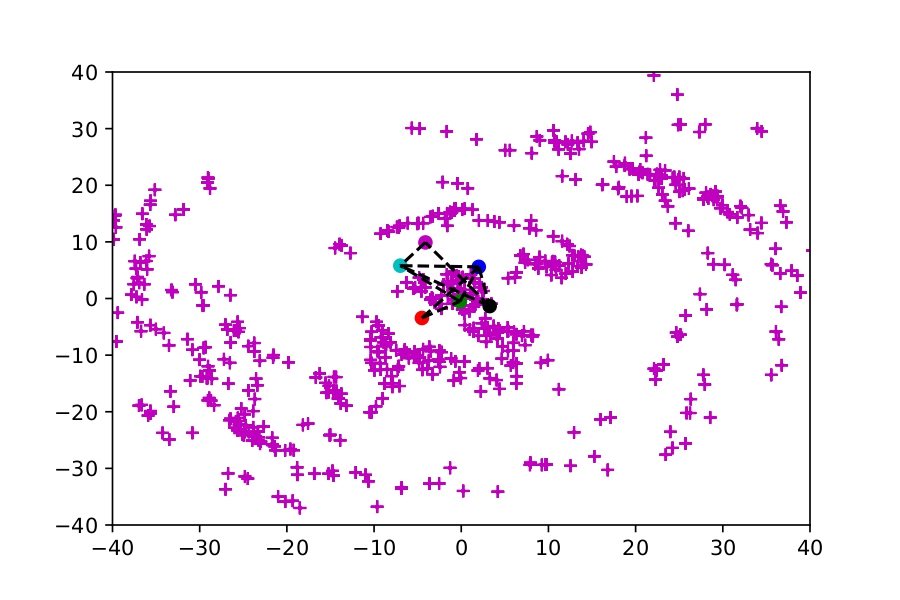}
\includegraphics[width=0.49\columnwidth]{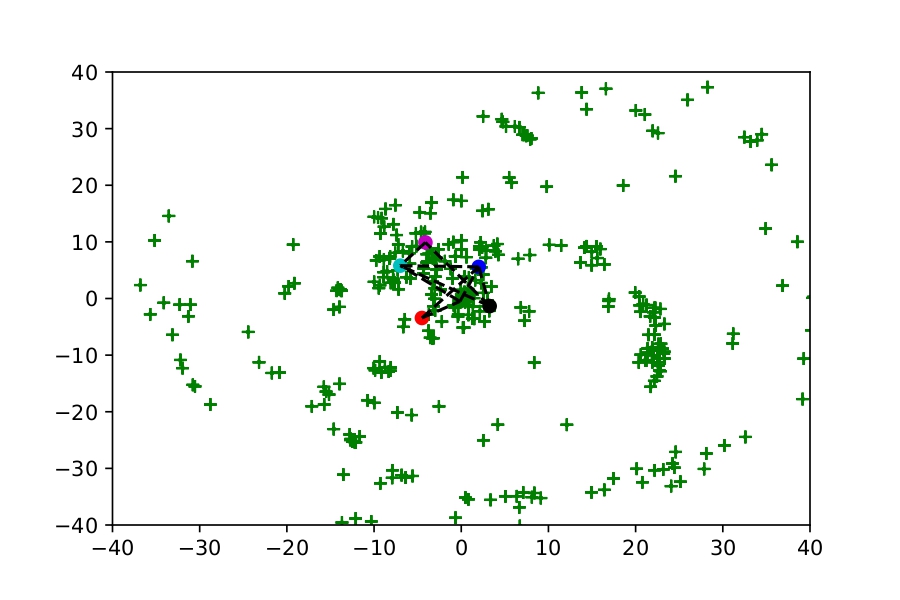}
\includegraphics[width=0.49\columnwidth]{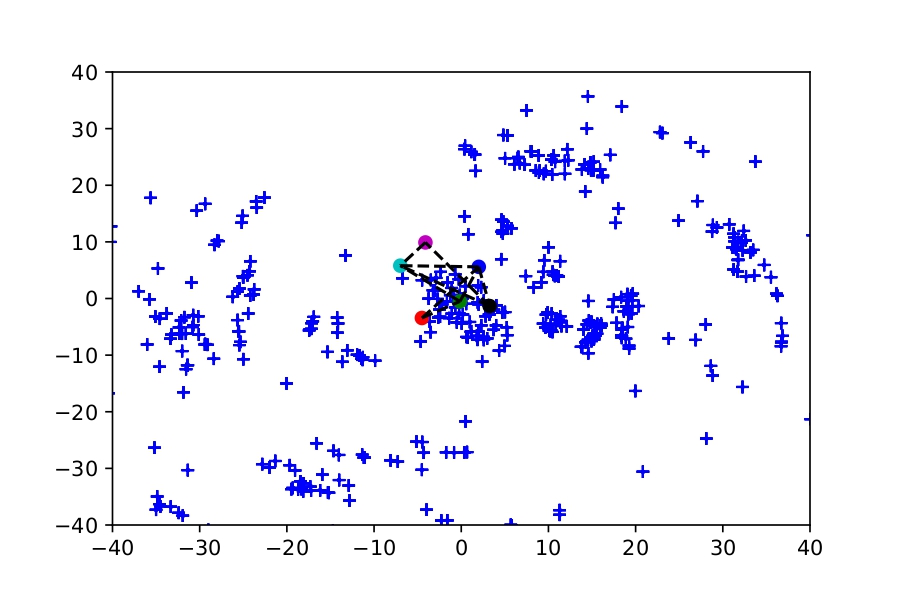}
	\caption{In these plots, all the agents except one keep all the desired distances in between at the beginning of the simulation. The variational integrator allows us to estimate the regions of attraction of the agent that has not been collocated correctly. Surprisingly, we identify that beyond the small perturbations of the desired position of the non-collocated agent, other areas form circular "halos" around the desired shape.}
\label{fig: sim3}
\end{figure}

We would like to highlight that the simulation campaign with the variational integrator takes around one hour per $3000$ simulations in an 	Intel(R) Core(TM) i7-8650U CPU. In this simulation campaign, the integration of the equations is the most expensive operation per iteration. Therefore, the proposed (explicit) variational integrator \eqref{integrator-explicit} assisted us in speeding up the time-consuming process.

\section{Conclusions}
\label{sec: con}

We have constructed variational integrators for non-autonomous Lagrangian systems with fixed time step. In particular, a variational integrator for a time-dependent Lagrangian system was derived via a family of discrete Lagrangian functions each one for a fixed time-step. This allows to recover at each step on the set of discrete sequences the preservation properties of variational integrators for autonomous Lagrangian systems such as symplecticity of the integrator or exponential decay of the energy due to backward error analysis. By assuming a regularity condition we can derive the corresponding discrete Hamiltonian flow. A Noether theorem for this class of systems was also obtained giving rise to a relation between noether symmetries and constants of the motion for both the continuous-time and the discrete-time Euler-Lagrange equations.  
In a further work we would like to study the applicability of backward error analysis in the Lagrangian side as in \cite{vre} but in the non-autonomous case and compare with the results obtained in this paper. 
Another perspective is the extension to time-dependent forced systems and applications to formation control \cite{CoGa3}.
\section*{Acknowledgments}
The authors acknowledge financial support from the Spanish Ministry of Science and Innovation, under grants PID2019-
106715GB-C21, MTM2016-76702-P, the ``Severo Ochoa Programme for Centres of Excellence''in R$\&$D (CEX2019-000904-S). This work was supported by a 2020 Leonardo Grant
for Researchers and Cultural Creators, BBVA Foundation. The BBVA Foundation accepts no responsibility
for the opinions, statements and contents included in the
project and/or the results thereof, which are entirely the
responsibility of the authors.

\end{document}